\documentclass[twoside,journal]{IEEEtran}
\ifCLASSINFOpdf
\else
    \usepackage[dvips]{graphicx}
\fi
\usepackage{amsmath,amssymb}
\newcommand{\bysame}{%
    \leavevmode\hbox to 3em{\hrulefill}\,}

\begin{document}
%
\title{An Innovations Approach to Viterbi Decoding of Convolutional Codes}
%
%
%

\author{Masato~Tajima,~\IEEEmembership{Senior~Member,~IEEE}
\thanks{Manuscript received October 30, 2017; revised May 14, 2018.}
\thanks{M. Tajima was with the graduate School of
Science and Engineering, University of Toyama, 3190 Gofuku,
Toyama 930-8555, Japan (e-mail: masatotjm@kind.ocn.ne.jp).}}

%
%

\markboth{IEEE Transactions on Information Theory,~Vol.~64, No.~9, September~2018}
{Tajima \MakeLowercase{\textit{}}: An Innovations Approach to Viterbi Decoding of Convolutional Codes}
%



\maketitle

\begin{abstract}
We introduce the notion of innovations for Viterbi decoding of convolutional codes. First we define a kind of innovation corresponding to the received data, i.e., the input to a Viterbi decoder. Then the structure of a Scarce-State-Transition (SST) Viterbi decoder is derived in a natural manner. It is shown that the newly defined innovation is just the input to the main decoder in an SST Viterbi decoder and generates the same syndrome as the original received data does. A similar result holds for Quick-Look-In (QLI) codes as well. In this case, however, the precise innovation is not defined. We see that this innovation-like quantity is related to the linear smoothed estimate of the information. The essence of innovations approach to a linear filtering problem is first to whiten the observed data, and then to treat the resulting simpler white-noise observations problem. In our case, this corresponds to the reduction of decoding complexity in the main decoder in an SST Viterbi decoder. We show the distributions related to the main decoder (i.e., the input distribution and the state distribution in the code trellis for the main decoder) are much biased under moderately noisy conditions. We see that these biased distributions actually lead to the complexity reduction in the main decoder. Furthermore, it is shown that the proposed innovations approach can be extended to maximum-likelihood (ML) decoding of block codes as well.
\end{abstract}


%
\IEEEpeerreviewmaketitle

\section{Introduction}
%
%
%
%
\begin{figure}[htb]
\begin{center}
\includegraphics[width=9.0cm,clip]{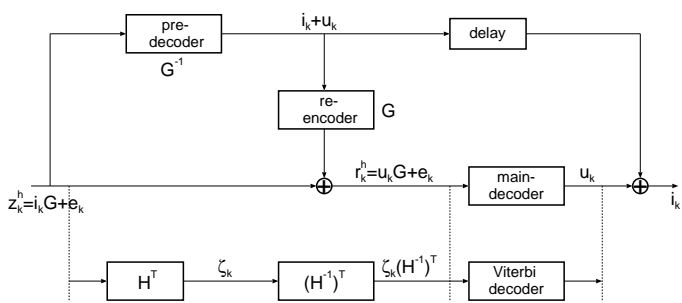}
\end{center}
\caption{The structure of an SST Viterbi decoder (pre-decoder: $G^{-1}$).}
\label{Fig.1}
\end{figure}
\IEEEPARstart{I}{n} 1985, Kubota, Kohri, and Kato~\cite{kubo 85} proposed a Viterbi decoding scheme named Scarce-State-Transition (SST) for the purpose of decoding of Quick-Look-In (QLI) codes~\cite{mass 71}. They also extended the scheme to general codes. The corresponding Viterbi decoder consists of a pre-decoder and a main decoder (i.e., a conventional Viterbi decoder). The structure of an SST Viterbi decoder is shown in Fig.1~\cite{taji 032}, where the inverse encoder is used as a pre-decoder. At the first stage, the transmitted information is estimated using a rather simple decoder (i.e., a pre-decoder) such as the inverse encoder, and then at the second stage, the estimation error at the first stage is decoded using a main decoder. Finally, two decoder outputs are combined to produce the final decoder output. The SST scheme was devised mainly for the purpose of hardware and power-consumption reduction in Viterbi decoder VLSI implementation. More precisely~\cite{kubo 86,kubo 93},
\begin{itemize}
\item [1)] A likelihood concentration to the all-zero state\footnote{The state in the code trellis for the main decoder consists of errors and is regarded as a discrete random variable. We call its distribution simply a {\it state distribution}. Then a {\it likelihood concentration} means that the state distribution is not uniform but {\it biased}.} occurs in the main decoder.
\item [2)] In the main decoder, a maximum-likelihood decision circuit, which is used to determine the most likely survivor from among all survivors at each depth, is omitted within a very small performance degradation.
\item [3)] On-off switching rarely occurs in the path-memory circuit in the main decoder when a decoder LSI is implemented using the CMOS technology.
\end{itemize}
\par
Since the estimation ``error'' is decoded in the main decoder, it is natural to think that the SST scheme is closely related to syndrome decoding~\cite{arie 95,arie 98,arie 99, scha 75,scha 76,scha 78} based on an error trellis. Later~\cite{taji 031,taji 032}, we showed that SST Viterbi decoding based on a code trellis and syndrome decoding based on the corresponding error trellis are equivalent under a general condition.
\par
On the other hand, in connection with stochastic processes, the problem of extracting the innovations~\cite{abe 12,hida 11,hida 14,kai 98,kuni 76,wong 73} from a given (complex) process has been discussed for a long time (see~\cite{hida 11,hida 14}). Let $X(t)$ be a stochastic process. Suppose that during an infinitesimal interval $[t, t+dt)$, $X(t)$ obtains new information which is independent of the information obtained by $X(t)$ prior to time $t$. The newly obtained information is called the ``innovation'' associated with $X(t)$. Kailath~\cite{kai 681} applied the notion of innovations to a linear filtering problem~\cite{ari 77,jaz 70,kai 681,kuni 76,saka 72,wong 71}. Also, Kailath and Frost~\cite{kai 682} extended the idea to a linear smoothing problem~\cite{jaz 70,kai 682,saka 72}. In the linear filtering theory, the innovation associated with an observation is defined by the difference between the observation and the estimate of a signal, or equivalently, the sum of the estimation error and a noise~\cite{kai 681,kai 682}. Hence, we thought the notion of innovations has some connection with SST Viterbi decoding in the coding theory.
\par
In this paper, by comparing with the results in the linear filtering theory, we define a kind of innovation corresponding to the received data for a Viterbi decoder. Then the structure of an SST Viterbi decoder is derived in a natural manner. We see that the newly defined innovation is just the input to the main decoder in an SST Viterbi decoder. A similar result is obtained in connection with QLI codes as well. In the latter case, however, the precise innovation is not defined. It is shown that the obtained innovation-like quantity is related to linear smoothing of the information. Moreover, for a QLI code, we examine the relationship between the two estimates of the information, i.e., the linear filtered estimate and the linear smoothed estimate. Then it is shown that the latter has higher accuracy as compared with the former. These are discussed in Section II.
\par
Now the main purpose of introducing the innovations in a filtering problem is to whiten the observed data~\cite{kai 681}. As a result, the given problem is transformed to a {\it simpler} white-noise observations problem. We thought this corresponds to the reduction of decoding complexity in the main decoder in an SST Viterbi decoder. The reduction of hardware and power-consumption of an LSI is also considered as a related simplification. Then we thought all of these reductions are caused by {\it biased} distributions related to the main decoder. Hence, in Section III, we focus our arguments mainly on these distributions. We see that the distribution of the input to the main decoder is biased under moderately noisy conditions. The state distribution in the code trellis for the main decoder is also biased under the same channel conditions. Moreover, we observe that the state distribution in the error trellis is equally biased.
\par
Subsequently, in Section IV, we show those biased distributions actually lead to the reduction of decoding complexity in the main decoder. Since there have been several related works~\cite{arie 95,arie 99,pin 91,taji 86,taji 971,taji 972}, the discussion is mainly based on these known works. We remark that syndrome decoding based on an error trellis has less complexity as compared with Viterbi decoding based on a code trellis~\cite{arie 95,arie 99}. Since the SST scheme is equivalent to syndrome decoding based on the error trellis, this is quite reasonable. In connection with the subject, we derive an approximate criterion for complexity reduction in the main decoder.
\par
The fundamental feature of the SST scheme lies in its structure where an {\it estimation error} is decoded in the main decoder. Then we see that a similar scheme (i.e., two-stage decoding) can be applied to block codes as well. In Section V, it is shown that a kind of innovation can also be extracted in connection with maximum-likelihood (ML) decoding of block codes~\cite{lin 04}. 
\par
Let us close this section by introducing the basic notions needed for this paper. We always assume that the underlying field is $\mbox{GF}(2)$. Let $G(D)$ be a generator matrix for an $(n_0, k_0)$ convolutional code, where $G(D)$ is assumed to be {\it canonical}~\cite{joha 99,mc 962} (i.e., {\it minimal}~\cite{forn 70}). A corresponding check matrix $H(D)$ is also assumed to be canonical. Hence, they have the same constraint length, denoted $\nu$. Denote by $\mbox{\boldmath $i$}\!=\!\{\mbox{\boldmath $i$}_k\}$ and $\mbox{\boldmath $y$}\!=\!\{\mbox{\boldmath $y$}_k\}$ an information sequence and the corresponding code sequence, respectively, where $\mbox{\boldmath $i$}_k\!=\!(i_k^{(1)}, \cdots, i_k^{(k_0)})$ is the information block at $t\!=\!k$ and $\mbox{\boldmath $y$}_k\!=\!(y_k^{(1)}, \cdots, y_k^{(n_0)})$ is the encoded block at $t\!=\!k$. In this paper, it is assumed that a code sequence {\boldmath $y$} is transmitted symbol by symbol over a memoryless AWGN channel using BPSK modulation~\cite{hell 71}. Let $\mbox{\boldmath $z$}\!=\!\{\mbox{\boldmath $z$}_k\}$ be a received sequence, where $\mbox{\boldmath $z$}_k\!=\!(z_k^{(1)}, \cdots, z_k^{(n_0)})$ is the received block at $t\!=\!k$. Each component $z_j$ of {\boldmath $z$} is modeled as
\begin{equation}
z_j=x_j\sqrt{2E_s/N_0}+w_j .
\end{equation}
Here, $x_j$ takes $\pm 1$ depending on whether the code symbol $y_j$ is $0$ or $1$. $E_s$ and $N_0$ denote the energy per channel symbol and the single-sided noise spectral density, respectively. (Let $E_b$ be the energy per information bit. Then the relationship between $E_b$ and $E_s$ is defined by $E_s\!=\!RE_b$, where $R$ is the code rate.) Also, $w_j$ is a zero-mean unit variance Gaussian random variable with probability density function
\begin{equation}
q(y)=\frac{1}{\sqrt{2\pi}}e^{-\frac{y^2}{2}} .
\end{equation}
Each $w_j$ is independent of all others. Let $p(z_j \vert y_j)$ be the conditional probability density function of $z_j$ given $y_j$. The hard-decision (denoted ``$^h$'') data of $z_j$ is defined by
\begin{equation}
z_j^h\stackrel{\triangle}{=}\left\{
\begin{array}{rl}
0,& \quad L(z_j \vert y_j) \geq 0 \\
1,& \quad L(z_j \vert y_j) < 0 ,
\end{array} \right.
\end{equation}
where
\begin{equation}
L(z_j \vert y_j)\stackrel{\triangle}{=}\log \frac{p(z_j \vert y_j=0)}{p(z_j \vert y_j=1)}
\end{equation}
is the log-likelihood ratio conditioned on $y_j$ (``$\log$'' denotes the natural logarithm). In our case, this is equivalent to
\begin{equation}
z_j^h\stackrel{\triangle}{=}\left\{
\begin{array}{rl}
0,& \quad z_j \geq 0 \\
1,& \quad z_j < 0 .
\end{array} \right.
\end{equation}
Note that in Fig.1, the main decoder input $r_k^{(l)}~(1 \leq l \leq n_0)$ is given by
\begin{equation}
r_k^{(l)}=\left\{
\begin{array}{rl}
\vert z_k^{(l)} \vert,& \quad r_k^{(l)h}=0 \\
- \vert z_k^{(l)} \vert,& \quad r_k^{(l)h}=1 .
\end{array} \right.
\end{equation}
\par
Let $\mbox{\boldmath $v$}_k\!=\!(v_k^1, \cdots, v_k^n)$ be an $n$-tuple of variables. Also, let $\mbox{\boldmath $p$}(D)\!=\!(p_1(D), \cdots, p_n(D))$ be an $n$-tuple of polynomials in $D$. Since each $p_i(D)$ is a delay operator with respect to $k$, $\sum_{i=1}^np_i(D)v_k^i$ is well defined, where $D^mv_k^i\!=\!v_{k-m}^i$. In this paper, noting that $\mbox{\boldmath $v$}_k$ is a row vector, we express the above variable as $\mbox{\boldmath $v$}_k\mbox{\boldmath $p$}^T(D)$ (``$^T$'' means transpose). Using this notation, we have
\begin{equation}
\mbox{\boldmath $y$}_k=\mbox{\boldmath $i$}_kG(D) .
\end{equation}
Also, the syndrome at $t\!=\!k$ is defined by
\begin{equation}
\mbox{\boldmath $\zeta$}_k=\mbox{\boldmath $z$}_k^hH^T(D) .
\end{equation}
Note that $\mbox{\boldmath $\zeta$}_k\!=\!\mbox{\boldmath $e$}_kH^T(D)$ holds, where $\mbox{\boldmath $e$}_k\!=\!(e_k^{(1)}, \cdots, e_k^{(n_0)})$ is the error at $t\!=\!k$.

\section{An Innovations Approach to Viterbi Decoding of Convolutional Codes}
As stated in the preceding section, it seems that the notion of innovations introduced for linear filtering/smoothing problems has some connection with SST Viterbi decoding of convolutional codes. In the following, based on this conjecture, we investigate Viterbi decoding of convolutional code from an innovation viewpoint.

\subsection{Innovations Associated with the Received Data for a Viterbi Decoder}
First consider a linear filtering problem~\cite{ari 77,jaz 70,kai 681,kuni 76,saka 72,wong 71}. Let
\begin{equation}
y(t)=C(t)x(t)+w(t)
\end{equation}
be the observation corresponding to a signal $x(t)$, where $C(t)$ is a coefficient matrix and $w(t)$ is a white Gaussian noise. In this case, the innovation $\nu(t)$~\cite{kai 681} associated with $y(t)$ is defined as
\begin{equation}
\nu(t)=y(t)-C(t)\hat x(t \vert t) ,
\end{equation}
where $\hat x(t \vert t)$ is a linear function of all the data $\{y(s), s<t\}$ that minimizes the mean-square error $E[(x(t)-\hat x(t \vert t))^T(x(t)-\hat x(t \vert t))]$ (``$E[\cdot]$'' is the expectation)~\cite{kai 681}.
\par
Next, consider convolutional encoding based on $G(D)$. Let 
\begin{equation}
\mbox{\boldmath $z$}_k^h=\mbox{\boldmath $i$}_kG(D)+\mbox{\boldmath $e$}_k
\end{equation}
be the received data, where $\mbox{\boldmath $i$}_k$ and $\mbox{\boldmath $e$}_k$ are an information block and an error, respectively. By comparison with the linear filtering theory, it is reasonable to think that
\begin{eqnarray}
\mbox{\boldmath $r$}_k^h &=& \mbox{\boldmath $z$}_k^h-\mbox{\boldmath $\hat i$}(k \vert k)G(D) \nonumber \\
&=& \mbox{\boldmath $z$}_k^h+\mbox{\boldmath $\hat i$}(k \vert k)G(D)
\end{eqnarray}
corresponds to $\nu(t)$, where $\mbox{\boldmath $\hat i$}(k \vert k)$ denotes an estimate of $\mbox{\boldmath $i$}_k$ based on $\{\mbox{\boldmath $z$}_s^h, s \leq k\}$. Suppose that $\mbox{\boldmath $\hat i$}(k \vert k)$ is a linear combination of the received data $\{\mbox{\boldmath $z$}_s^h, s \leq k\}$ and has the form
\begin{equation}
\mbox{\boldmath $\hat i$}(k \vert k)=\mbox{\boldmath $z$}_k^hP(D) ,
\end{equation}
where $P(D)$ is a polynomial matrix. Then we have
\begin{eqnarray}
\mbox{\boldmath $r$}_k^h &=& \mbox{\boldmath $z$}_k^h+\mbox{\boldmath $z$}_k^hP(D)G(D) \nonumber \\
&=& (\mbox{\boldmath $i$}_kG(D)+\mbox{\boldmath $e$}_k)+(\mbox{\boldmath $i$}_kG(D)+\mbox{\boldmath $e$}_k)P(D)G(D) \nonumber \\
&=& \mbox{\boldmath $i$}_k(I_{k_0}\!+\!G(D)P(D))G(D)\!+\!\mbox{\boldmath $e$}_kP(D)G(D)\!+\!\mbox{\boldmath $e$}_k , \nonumber
\end{eqnarray}
where $I_{k_0}$ is the identity matrix of size $k_0 \times k_0$. Note that if
\begin{displaymath}
(I_{k_0}\!+\!G(D)P(D))G(D)\!=\!G(D)\!+\!G(D)P(D)G(D)\!=\!0
\end{displaymath}
or
\begin{equation}
G(D)P(D)G(D)=G(D)
\end{equation}
holds, then $\mbox{\boldmath $r$}_k^h$ is independent of $\mbox{\boldmath $i$}_k$. Here $G(D)P(D)G(D)\!=\!G(D)$ implies that $P(D)$ is a {\it generalized inverse}~\cite{rao 72} of $G(D)$. Then a right inverse $G^{-1}(D)$ of $G(D)$ can be taken as $P(D)$. In this case, $\mbox{\boldmath $r$}_k^h$ is independent of $\mbox{\boldmath $i$}_k$ and we have
\begin{eqnarray}
\mbox{\boldmath $r$}_k^h &=& (\mbox{\boldmath $e$}_kG^{-1})G+\mbox{\boldmath $e$}_k \\
&=& \mbox{\boldmath $u$}_kG+\mbox{\boldmath $e$}_k \\
&=& \mbox{\boldmath $e$}_k(G^{-1}G+I_{n_0}) ,
\end{eqnarray}
where $\mbox{\boldmath $u$}_k\stackrel{\triangle}{=}\mbox{\boldmath $e$}_kG^{-1}$. We think this quantity corresponds to an innovation in the linear filtering theory. We remark that the right-hand side is just the input to the main decoder in an SST Viterbi decoder, where the inverse encoder $G^{-1}$ is used as a pre-decoder (see Fig.1). Also, note that
\begin{eqnarray}
\mbox{\boldmath $r$}_k^hH^T(D) &=& \mbox{\boldmath $z$}_k^hH^T(D)+\mbox{\boldmath $z$}_k^hP(D)G(D)H^T(D) \nonumber \\
&=& \mbox{\boldmath $z$}_k^hH^T(D)=\mbox{\boldmath $\zeta$}_k
\end{eqnarray}
holds irrespective of $P(D)$. Hence, $\mbox{\boldmath $r$}_k^h$ and $\mbox{\boldmath $z$}_k^h$ generate the same syndrome $\mbox{\boldmath $\zeta$}_k$.
\par
On the other hand, $\mbox{\boldmath $r$}_k^h$ has another expression. Let
\begin{equation}
G=A \times \Gamma \times B
\end{equation}
be an invariant-factor decomposition~\cite{forn 70} of $G(D)$. Since $G(D)$ is canonical (accordingly, basic), we can assume~\cite{forn 70} that the first $k_0$ rows of $B$ coincide with $G(D)$ and the last $(n_0-k_0)$ columns of $B^{-1}$ coincide with the syndrome former $H^T(D)$. As a result, we have
\begin{eqnarray}
I_{n_0} &=& B^{-1}B \nonumber \\
 &=& \left(
\begin{array}{cc}
G^{-1} & H^T 
\end{array}
\right)\left(
\begin{array}{c}
G \\
(H^{-1})^T
\end{array}
\right) \nonumber \\
&=& G^{-1}G+H^T(H^{-1})^T .
\end{eqnarray}
Then
\begin{eqnarray}
\mbox{\boldmath $r$}_k^h &=& \mbox{\boldmath $e$}_k(G^{-1}G+I_{n_0}) \nonumber \\
&=& \mbox{\boldmath $e$}_kH^T(H^{-1})^T=\mbox{\boldmath $\zeta$}_k(H^{-1})^T
\end{eqnarray}
is obtained. Thus we have again
\begin{equation}
\mbox{\boldmath $r$}_k^hH^T=\mbox{\boldmath $\zeta$}_k(H^{-1})^TH^T=\mbox{\boldmath $\zeta$}_k . \nonumber
\end{equation}
\par
Therefore, $\mbox{\boldmath $r$}_k^h$ has the following properties:
\begin{itemize}
\item[1)] $\mbox{\boldmath $r$}_k^h\!=\!\mbox{\boldmath $e$}_k(G^{-1}G\!+\!I_{n_0})$ holds. Hence, $\mbox{\boldmath $r$}_k^h$ consists of errors $\{\mbox{\boldmath $e$}_s,~s \leq k\}$. There is a correspondence between $\mbox{\boldmath $e$}_k$ and $\mbox{\boldmath $r$}_k^h$ in the sense that they generate the same syndrome $\mbox{\boldmath $\zeta$}_k$.
\item[2)] $\{\mbox{\boldmath $r$}_s^h,~s \leq k\}$ and $\{\mbox{\boldmath $z$}_s^h,~s \leq k\}$ generate the same syndrome sequence $\{\mbox{\boldmath $\zeta$}_s,~s \leq k\}$.
\end{itemize}
Property 1) corresponds to the fact that an innovation process is a white-noise process in the linear filtering theory. Property 2) is the most important one and corresponds to the fact that the original received data and the associated innovations have the same information. In the case of error correction, if two quantities generate the same syndrome sequence, then we can conclude that they have the equal information. Here we remark that $\{\mbox{\boldmath $r$}_k^h\}$ does not have the same properties as those of innovations in the linear filtering theory. Hence, we may call $\{\mbox{\boldmath $r$}_k^h\}$ the innovations associated with $\{\mbox{\boldmath $z$}_k^h\}$ in a {\it weak sense}~\cite{kuni 76}. All of this leads to the following notation.
\newtheorem{df}{Definition}[section]
\begin{df}
Let $\{\mbox{\boldmath $z$}_k^h\}$ be the received data. Here assume the following: For $\mbox{\boldmath $z$}_k^h$, there exists $\mbox{\boldmath $r$}_k^h$ which consists of errors $\{\mbox{\boldmath $e$}_s,~s \leq k\}$ such that for each $k$, $\{\mbox{\boldmath $r$}_s^h,~s \leq k\}$ and $\{\mbox{\boldmath $z$}_s^h,~s \leq k\}$ generate the same syndrome sequence $\{\mbox{\boldmath $\zeta$}_s,~s \leq k\}$. In this case, we call $\{\mbox{\boldmath $r$}_k^h\}$ the innovations associated with $\{\mbox{\boldmath $z$}_k^h\}$.
\end{df}
\par
The above argument implies that we may call
\begin{eqnarray}
\mbox{\boldmath $r$}_k^h &=& \mbox{\boldmath $z$}_k^h+(\mbox{\boldmath $z$}_k^hG^{-1})G \nonumber \\
&=& \mbox{\boldmath $z$}_k^h(I_{n_0}+G^{-1}G)
\end{eqnarray}
the {\it innovation} corresponding to $\mbox{\boldmath $z$}_k^h$.
\par
Here note the mapping: $\mbox{\boldmath $z$}_k^h \mapsto \mbox{\boldmath $r$}_k^h$. In the innovations approach to linear filtering problems, the observed data is whitened by a causal~\cite{forn 70} and invertible operation. With respect to the above mapping, we have the following.
\newtheorem{pro}[df]{Proposition}
\begin{pro}
The mapping: $\mbox{\boldmath $z$}_k^h \mapsto \mbox{\boldmath $r$}_k^h\!=\!\mbox{\boldmath $z$}_k^h(I_{n_0}\!+\!G^{-1}G)$ is not invertible.
\end{pro}
\begin{IEEEproof}
We will show that $\mbox{det}(I_{n_0}\!+\!G^{-1}G)\!=\!\mbox{det}\bigl(H^T(H^{-1})^T \bigr)\!=\!0$ (``$\mbox{det}(\cdot)$'' is the determinant). Since $H$ is assumed to be canonical (accordingly, basic), we have a following invariant-factor decomposition:
\begin{displaymath}
H=\hat A \times \hat \Gamma \times \hat B ,
\end{displaymath}
where
\begin{eqnarray}
\hat \Gamma &=& \left(
\begin{array}{cccc|ccc}
1 & 0 & \scriptstyle{\ldots} & 0 & 0 & \scriptstyle{\ldots} & 0 \\
0 & 1 & \scriptstyle{\ldots} & 0 & 0 & \scriptstyle{\ldots} & 0 \\
\scriptstyle{\vdots} & \scriptstyle{\vdots} & \scriptstyle{\ddots} & \scriptstyle{\vdots} & \scriptstyle{\vdots} & \scriptstyle{\ddots} & \scriptstyle{\vdots} \\
0 & 0 & \scriptstyle{\ldots} & 1 & 0 & \scriptstyle{\ldots} & 0 
\end{array}
\right) \nonumber \\
&\stackrel{\triangle}{=}& \left(
\begin{array}{cc}
I_{n_0-k_0} & O_{n_0-k_0, k_0}
\end{array}
\right) . \nonumber
\end{eqnarray}
Here, $O_{n_0-k_0, k_0}$ denotes the zero matrix of size $(n_0-k_0) \times k_0$. Then~\cite{joha 99} we have
\begin{displaymath}
H^{-1}=\hat B^{-1} \times \hat \Gamma^{-1} \times \hat A^{-1} ,
\end{displaymath}
where
\begin{eqnarray}
\hat \Gamma^{-1} &=& \left(
\begin{array}{cccc}
1 & 0 & \scriptstyle{\ldots} & 0 \\
0 & 1 & \scriptstyle{\ldots} & 0 \\
\scriptstyle{\vdots}& \scriptstyle{\vdots}& \scriptstyle{\ddots} & \scriptstyle{\vdots} \\
0 & 0 & \scriptstyle{\ldots} & 1 \\
\hline
0 & 0 & \scriptstyle{\ldots} & 0 \\
\scriptstyle{\vdots}& \scriptstyle{\vdots}& \scriptstyle{\ddots} & \scriptstyle{\vdots} \\
0 & 0 & \scriptstyle{\ldots} & 0
\end{array}
\right) \nonumber \\
&\stackrel{\triangle}{=}& \left(
\begin{array}{c}
I_{n_0-k_0} \\
O_{k_0, n_0-k_0}
\end{array}
\right) . \nonumber
\end{eqnarray}
Hence, it follows that
\begin{eqnarray}
H^{-1}H &=& \hat B^{-1}\hat \Gamma^{-1}\hat A^{-1}\hat A \hat \Gamma \hat B \nonumber \\
&=& \hat B^{-1}\hat \Gamma^{-1}\hat \Gamma \hat B \nonumber \\
&=& \hat B^{-1}\left(
\begin{array}{c}
I_{n_0-k_0} \\
O_{k_0, n_0-k_0}
\end{array}
\right) \nonumber \\
&& \qquad \times \left(
\begin{array}{cc}
I_{n_0-k_0} & O_{n_0-k_0, k_0}
\end{array}
\right)\hat B \nonumber \\
&=& \hat B^{-1}\left(
\begin{array}{cc}
I_{n_0-k_0} & O_{n_0-k_0, k_0} \\
O_{k_0, n_0-k_0} & O_{k_0, k_0}
\end{array}
\right)\hat B . \nonumber
\end{eqnarray}
Accordingly,
\begin{displaymath}
H^T(H^{-1})^T=\hat B^T\left(
\begin{array}{cc}
I_{n_0-k_0} & O_{n_0-k_0, k_0} \\
O_{k_0, n_0-k_0} & O_{k_0, k_0}
\end{array}
\right)(\hat B^{-1})^T .
\end{displaymath}
Hence, we have
\begin{eqnarray}
\mbox{det}\bigl(H^T(H^{-1})^T \bigr) &=& \mbox{det}\bigl(\hat B^T \bigr)\mbox{det}\left(
\begin{array}{cc}
I_{n_0-k_0} & O_{n_0-k_0, k_0} \\
O_{k_0, n_0-k_0} & O_{k_0, k_0}
\end{array}
\right) \nonumber \\
&& \times~\mbox{det}\bigl((\hat B^{-1})^T \bigr) \nonumber \\
&=& \mbox{det}(\hat B)\mbox{det}\left(
\begin{array}{cc}
I_{n_0-k_0} & O_{n_0-k_0, k_0} \\
O_{k_0, n_0-k_0} & O_{k_0, k_0}
\end{array}
\right) \nonumber \\
&& \times~\mbox{det}(\hat B^{-1}) \nonumber \\
&=& \mbox{det}\left(
\begin{array}{cc}
I_{n_0-k_0} & O_{n_0-k_0, k_0} \\
O_{k_0, n_0-k_0} & O_{k_0, k_0}
\end{array}
\right) . \nonumber
\end{eqnarray}
Finally, note that
\begin{displaymath}
\mbox{det}\left(
\begin{array}{cc}
I_{n_0-k_0} & O_{n_0-k_0, k_0} \\
O_{k_0, n_0-k_0} & O_{k_0, k_0}
\end{array}
\right)=0 .
\end{displaymath}
\end{IEEEproof}
\par
The following shows that the innovation $\mbox{\boldmath $r$}_k^h$ corresponding to $\mbox{\boldmath $z$}_k^h$ cannot be further reduced.
\begin{pro}
In the relation $\mbox{\boldmath $r$}_k^h\!=\!\mbox{\boldmath $z$}_k^h(I_{n_0}\!+\!G^{-1}G)$, replace $\mbox{\boldmath $z$}_k^h$ on the right-hand side by $\mbox{\boldmath $r$}_k^h$. Then we have $\mbox{\boldmath $r$}_k^h$ again.
\end{pro}
\begin{IEEEproof}
\begin{eqnarray}
\lefteqn{\mbox{\boldmath $r$}_k^h(I_{n_0}+G^{-1}G)} \nonumber \\
&& =\mbox{\boldmath $r$}_k^hH^T(H^{-1})^T \nonumber \\
&& =\mbox{\boldmath $\zeta$}_k(H^{-1})^T=\mbox{\boldmath $r$}_k^h .
\end{eqnarray}
\end{IEEEproof}

\subsection{Relationship Between General Codes and QLI Codes}
\begin{figure}[htb]
\begin{center}
\includegraphics[width=9.5cm,clip]{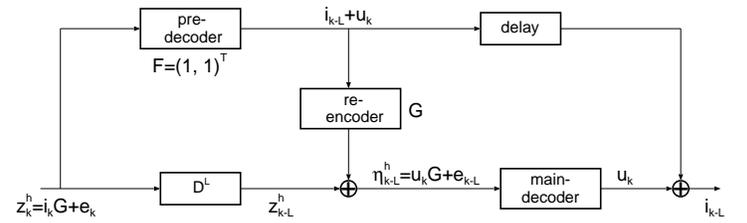}
\end{center}
\caption{The structure of an SST Viterbi decoder for a QLI code (pre-decoder: $F\!=\!(1, 1)^T$).}
\label{Fig.2}
\end{figure}
We remark that the first paper~\cite{kubo 85} on SST Viterbi decoding dealt with QLI codes. Let
\begin{equation}
G(D)=(g_1(D), g_2(D))
\end{equation}
\begin{displaymath}
(g_1+g_2=D^L,~1 \leq L \leq \nu-1)
\end{displaymath}
be a generator matrix for a QLI code, where $\nu$ is the constraint length of $G(D)$. The corresponding SST Viterbi decoder is shown in Fig.2~\cite{taji 032}.
\par
Here consider the following quantity:
\begin{eqnarray}
\mbox{\boldmath $\eta$}_{k-L}^h &=& \mbox{\boldmath $z$}_{k-L}^h-\hat i(k-L \vert k)G(D) \nonumber \\
&=& \mbox{\boldmath $z$}_{k-L}^h+\hat i(k-L \vert k)G(D) ,
\end{eqnarray}
where $\hat i(k-L \vert k)$ denotes an estimate of $i_{k-L}$ based on $\{\mbox{\boldmath $z$}_s^h, s \leq k\}$. In the linear filtering/smoothing theory, this corresponds to
\begin{equation}
y(t)-C(t)\hat x(t \vert b)~(t<b) .
\end{equation}
Hence, $\mbox{\boldmath $\eta$}_{k-L}^h$ is slightly different from the innovation associated with the observation $\mbox{\boldmath $z$}_{k-L}^h$. We can call $\hat i(k-L \vert k)$ a linear {\it smoothed} estimate of $i_{k-L}$. Note that $\hat x(t \vert b)$ is the estimate of $x(t)~(t<b)$ based on the observations $y(s)~(s<b)$~\cite{kai 682}. That is, more observations are used for the estimation of $x(t)$ as compared with $\hat x(t \vert t)$. Accordingly, the accuracy of $\hat x(t \vert b)$ may increase as compared with $\hat x(t \vert t)$. Then it is reasonable to think a similar result holds with respect to $\hat i(k-L \vert k)$ (see Proposition 2.8).
\par
Now suppose that $\hat i(k-L \vert k)$ has the form
\begin{equation}
\hat i(k-L \vert k)=\mbox{\boldmath $z$}_k^hQ(D) ,
\end{equation}
where $Q(D)$ is a polynomial matrix. Then we have
\begin{eqnarray}
\mbox{\boldmath $\eta$}_{k-L}^h &=& \mbox{\boldmath $z$}_{k-L}^h+\mbox{\boldmath $z$}_k^hQ(D)G(D) \nonumber \\
&=& (i_{k-L}G(D)+\mbox{\boldmath $e$}_{k-L}) \nonumber \\
&& \qquad +(i_kG(D)+\mbox{\boldmath $e$}_k)Q(D)G(D) \nonumber \\
&=& i_k(D^L+G(D)Q(D))G(D) \nonumber \\
&& \qquad +\mbox{\boldmath $e$}_kQ(D)G(D)+\mbox{\boldmath $e$}_{k-L} . \nonumber
\end{eqnarray}
Note that if
\begin{displaymath}
(D^L\!+\!G(D)Q(D))G(D)\!=\!D^LG(D)\!+\!G(D)Q(D)G(D)\!=\!0
\end{displaymath}
or
\begin{equation}
G(D)D^{-L}Q(D)G(D)=G(D)
\end{equation}
holds, then $\mbox{\boldmath $\eta$}_{k-L}^h$ is independent of $i_k$. Here $G(D)D^{-L}Q(D)G(D)\!=\!G(D)$ implies that $D^{-L}Q(D)$ is a {\it generalized inverse}~\cite{rao 72} of $G(D)$. Then we can take $F\stackrel{\triangle}{=}\left(
\begin{array}{c}
1 \\
1
\end{array}
\right)$
as $Q(D)$. In this case, $\mbox{\boldmath $\eta$}_{k-L}^h$ is independent of $i_k$ and we have
\begin{eqnarray}
\mbox{\boldmath $\eta$}_{k-L}^h &=& (\mbox{\boldmath $e$}_kF)G+\mbox{\boldmath $e$}_{k-L} \\
&=& u_kG+\mbox{\boldmath $e$}_{k-L} \\
&=& \mbox{\boldmath $e$}_k(FG+D^LI_2) ,
\end{eqnarray}
where $u_k\stackrel{\triangle}{=}\mbox{\boldmath $e$}_kF$. We remark that the right-hand side is just the input to the main decoder in an SST Viterbi decoder, where $F$ is used as a pre-decoder (see Fig.2). Also, note that
\begin{eqnarray}
\mbox{\boldmath $\eta$}_{k-L}^hH^T(D) &=& \mbox{\boldmath $z$}_{k-L}^hH^T(D)\!+\!\mbox{\boldmath $z$}_k^hQ(D)G(D)H^T(D) \nonumber \\
&=& \mbox{\boldmath $z$}_{k-L}^hH^T(D)=\zeta_{k-L}
\end{eqnarray}
holds irrespective of $Q(D)$. Hence, $\mbox{\boldmath $\eta$}_{k-L}^h$ and $\mbox{\boldmath $z$}_{k-L}^h$ generate the same syndrome $\zeta_{k-L}$.
\par
On the other hand, $\mbox{\boldmath $\eta$}_{k-L}^h$ has another expression. We have
\begin{eqnarray}
FG+D^LI_2 &=& \left(
\begin{array}{cc}
g_1\!+\!D^L & g_2 \\
g_1 & g_2\!+\!D^L
\end{array}
\right) \nonumber \\
&=& \left(
\begin{array}{cc}
g_2 & g_2 \\
g_1 & g_1
\end{array}
\right) \nonumber \\
&=& (H^T, H^T) ,
\end{eqnarray}
where $H^T\!=\!\left(
\begin{array}{c}
g_2 \\
g_1
\end{array}
\right)$ is the syndrome former corresponding to $G\!=\!(g_1, g_2)$. Then
\begin{equation}
\mbox{\boldmath $\eta$}_{k-L}^h=\mbox{\boldmath $e$}_k(H^T, H^T)=(\zeta_k, \zeta_k)
\end{equation}
is obtained. Thus we have again
\begin{eqnarray}
\mbox{\boldmath $\eta$}_{k-L}^hH^T &=& (\zeta_k, \zeta_k)\left(
\begin{array}{c}
g_2 \\
g_1
\end{array}
\right) \nonumber \\
&=& \zeta_k(g_1+g_2) \nonumber \\
&=& \zeta_kD^L=\zeta_{k-L} . \nonumber
\end{eqnarray}
\par
Therefore, $\mbox{\boldmath $\eta$}_{k-L}^h$ has the following properties:
\begin{itemize}
\item[1)] $\mbox{\boldmath $\eta$}_{k-L}^h\!=\!\mbox{\boldmath $e$}_k(FG\!+\!D^LI_2)$ holds. Hence, $\mbox{\boldmath $\eta$}_{k-L}^h$ depends not only on errors $\{\mbox{\boldmath $e$}_s,~s \leq k-L\}$ but also on errors $\{\mbox{\boldmath $e$}_s,~k-L<s \leq k\}$ in general. There is a correspondence between $\mbox{\boldmath $e$}_k$ and $\mbox{\boldmath $\eta$}_{k-L}^h$ in the sense that the former generates the syndrome $\zeta_k$ and the latter generates the syndrome $\zeta_{k-L}$.
\item[2)] $\{\mbox{\boldmath $\eta$}_s^h,~s \leq k-L\}$ and $\{\mbox{\boldmath $z$}_s^h,~s \leq k-L\}$ generate the same syndrome sequence $\{\zeta_s,~s \leq k-L\}$.
\end{itemize}
\par
The above argument implies that
\begin{eqnarray}
\mbox{\boldmath $\eta$}_{k-L}^h &=& \mbox{\boldmath $z$}_{k-L}^h+(\mbox{\boldmath $z$}_k^hF)G \nonumber \\
&=& \mbox{\boldmath $z$}_k^h(D^LI_2+FG)
\end{eqnarray}
is not the innovation corresponding to $\mbox{\boldmath $z$}_{k-L}^h$ in the meaning of Definition 2.1.
\par
Now with respect to the mapping: $\mbox{\boldmath $z$}_k^h \mapsto \mbox{\boldmath $\eta$}_{k-L}^h$, we have the following.
\begin{pro}
The mapping: $\mbox{\boldmath $z$}_k^h \mapsto \mbox{\boldmath $\eta$}_{k-L}^h\!=\!\mbox{\boldmath $z$}_k^h(D^LI_2\!+\!FG)$ is not invertible.
\end{pro}
\begin{IEEEproof}
It follows from
\begin{displaymath}
D^LI_2+FG=\left(
\begin{array}{cc}
g_2 & g_2 \\
g_1 & g_1
\end{array}
\right)
\end{displaymath}
that $\mbox{det}(D^LI_2\!+\!FG)\!=\!0$.
\end{IEEEproof}
\par
The following shows that $\mbox{\boldmath $\eta$}_{k-L}^h$ cannot be further reduced as in the case of $\mbox{\boldmath $r$}_k^h$.
\begin{pro}
In the relation $\mbox{\boldmath $\eta$}_{k-L}^h\!=\!\mbox{\boldmath $z$}_k^h(D^LI_2\!+\!FG)$, replace $\mbox{\boldmath $z$}_k^h$ on the right-hand side by $\mbox{\boldmath $\eta$}_k^h$. Then we have $\mbox{\boldmath $\eta$}_{k-L}^h$ again.
\end{pro}
\begin{IEEEproof}
\begin{eqnarray}
\lefteqn{\mbox{\boldmath $\eta$}_k^h(D^LI_2+FG)} \nonumber \\
&& =\mbox{\boldmath $\eta$}_k^h(H^T, H^T) \nonumber \\
&& =(\zeta_k, \zeta_k)=\mbox{\boldmath $\eta$}_{k-L}^h .
\end{eqnarray}
\end{IEEEproof}
\par
Consider a QLI code defined by $G(D)$. It can be regarded as a general code as well. Hence, we can apply the argument in the preceding section to it. Let $\hat i(k-L \vert k)$ be the estimate of $i_{k-L}$ derived as a QLI code, whereas let $\hat i(k-L \vert k-L)$ be the estimate of $i_{k-L}$ derived as a general code. Then we have the following.
\begin{pro}
Let $G\!=\!(g_1, g_2)~(g_1\!+\!g_2\!=\!D^L)$ be a generator matrix for a QLI code. Define as follows:
\begin{eqnarray}
\hat i(k-L \vert k) &\stackrel{\triangle}{=}& \mbox{\boldmath $z$}_k^hF \\
\hat i(k-L \vert k-L) &\stackrel{\triangle}{=}& \mbox{\boldmath $z$}_{k-L}^hG^{-1} .
\end{eqnarray}
Then we have
\begin{equation}
\hat i(k-L \vert k)=\hat i(k-L \vert k-L)+\zeta_k ,
\end{equation}
where $\zeta_k\!=\!\mbox{\boldmath $e$}_kH^T\!=\!\mbox{\boldmath $e$}_k\left(
\begin{array}{c}
g_2 \\
g_1
\end{array}
\right)$ is the syndrome.
\end{pro}
\begin{IEEEproof}
From
\begin{eqnarray}
\hat i(k-L \vert k) &=& \mbox{\boldmath $z$}_k^hF=i_{k-L}+\mbox{\boldmath $e$}_kF \nonumber \\
\hat i(k-L \vert k-L) &=& \mbox{\boldmath $z$}_{k-L}^hG^{-1}\!=\!i_{k-L}\!+\!\mbox{\boldmath $e$}_{k-L}G^{-1} , \nonumber
\end{eqnarray}
the difference between $\hat i(k-L \vert k)$ and $\hat i(k-L \vert k-L)$ is given by
\begin{equation}
\mbox{\boldmath $e$}_kF+\mbox{\boldmath $e$}_{k-L}G^{-1}=\mbox{\boldmath $e$}_k(F+D^LG^{-1}) .
\end{equation}
Let
\begin{displaymath}
G^{-1}=\left(
\begin{array}{c}
b_1 \\
b_2
\end{array}
\right) .
\end{displaymath}
Then we have
\begin{equation}
F+D^LG^{-1}=\left(
\begin{array}{c}
1\!+\!D^Lb_1 \\
1\!+\!D^Lb_2
\end{array}
\right) .
\end{equation}
We show that the above is equal to $H^T$. In fact, we have
\begin{eqnarray}
\lefteqn{(g_1, g_2)\left(
\begin{array}{c}
1\!+\!D^Lb_1 \\
1\!+\!D^Lb_2
\end{array}
\right)} \nonumber \\
&& =(g_1+g_2)+D^L(g_1b_1+g_2b_2) \nonumber \\
&& =D^L+D^L=0 . \nonumber
\end{eqnarray}
\end{IEEEproof}
\par
\newtheorem{cor}[df]{Corollary}
\begin{cor}
Under the same conditions as in Proposition 2.6,
\begin{equation}
\mbox{\boldmath $\eta$}_{k-L}^h=\mbox{\boldmath $r$}_{k-L}^h+\zeta_kG
\end{equation}
holds.
\end{cor}
\begin{IEEEproof}
From
\begin{displaymath}
\mbox{\boldmath $z$}_k^hF=\mbox{\boldmath $z$}_{k-L}^hG^{-1}+\zeta_k ,
\end{displaymath}
it follows that
\begin{displaymath}
\mbox{\boldmath $z$}_{k-L}^h\!+\!(\mbox{\boldmath $z$}_k^hF)G=\mbox{\boldmath $z$}_{k-L}^h\!+\!(\mbox{\boldmath $z$}_{k-L}^hG^{-1})G\!+\!\zeta_kG .
\end{displaymath}
Here, it suffices to note the following equalities:
\begin{eqnarray}
\mbox{\boldmath $\eta$}_{k-L}^h &=& \mbox{\boldmath $z$}_{k-L}^h+(\mbox{\boldmath $z$}_k^hF)G \nonumber \\
\mbox{\boldmath $r$}_{k-L}^h &=& \mbox{\boldmath $z$}_{k-L}^h+(\mbox{\boldmath $z$}_{k-L}^hG^{-1})G . \nonumber
\end{eqnarray}
\end{IEEEproof}
\par
On the analogy of the linear filtering/smoothing theory, it is expected that the linear smoothed estimate $\hat i(k-L \vert k)$ has higher accuracy as compared with the linear filtered estimate $\hat i(k-L \vert k-L)$. In the following, $P(\cdot)$ denotes the probability and
\begin{equation}
\epsilon=\frac{1}{\sqrt{2\pi}} \int_{\sqrt{2E_s/N_0}}^{\infty}e^{-\frac{y^2}{2}}dy \stackrel{\triangle}{=}Q(\sqrt{2E_s/N_0})
\end{equation}
is the channel error probability. We have the following.
\begin{pro}
Let
\begin{eqnarray}
p_f &\stackrel{\triangle}{=}& P(\hat i(k-L \vert k-L) \neq i_{k-L}) \nonumber \\
&=& P(\mbox{\boldmath $e$}_{k-L}G^{-1}=1) \\
p_s &\stackrel{\triangle}{=}& P(\hat i(k-L \vert k) \neq i_{k-L}) \nonumber \\
&=& P(\mbox{\boldmath $e$}_kF=1) .
\end{eqnarray}
Then $p_s \leq p_f$ for $0 \leq \epsilon \leq 1/2$.
\end{pro}
\begin{IEEEproof}
Let $G^{-1}\!=\!\left(
\begin{array}{c}
b_1 \\
b_2
\end{array}
\right)$. Then $\mbox{\boldmath $e$}_{k-L}G^{-1}\!=\!1$ is expressed as
\begin{displaymath}
e_{k-L}^{(1)}b_1(D)+e_{k-L}^{(2)}b_2(D)=1 .
\end{displaymath}
We can rewrite the above as $e_1\!+\!e_2\!+\!\cdots \!+\!e_m\!=\!1$, where errors $e_j~(1 \leq j \leq m, 3 \leq m)$ are statistically independent of each other. Also, note that under this condition,
\begin{displaymath}
P \bigl(e_k^{(1)}\!+\!e_k^{(2)}\!=\!1 \bigr)=P(e_1\!+\!e_2\!=\!1)
\end{displaymath} 
holds. Hence, the comparison between $p_f$ and $p_s$ is reduced to that between $P(e_1\!+\!e_2\!+\!\cdots \!+\!e_m\!=\!1)$ and $P(e_1\!+\!e_2\!=\!1)$. Now we have
\begin{eqnarray}
\lefteqn{P(e_1\!+\!e_2\!+\!\cdots \!+\!e_m\!=\!1)} \nonumber \\
&& =P(e_1\!+\!e_2\!+\!e_b\!=\!1) \nonumber \\
&& =P(e_1\!+\!e_2\!=\!1, e_b\!=\!0)+P(e_1\!+\!e_2\!=\!0, e_b\!=\!1) \nonumber \\
&& =P(e_1\!+\!e_2\!=\!1)P(e_b\!=\!0)+P(e_1\!+\!e_2\!=\!0)P(e_b\!=\!1) , \nonumber
\end{eqnarray}
where $e_b\stackrel{\triangle}{=}e_3\!+\!\cdots \!+\!e_m$. Hence, we have
\begin{eqnarray}
\lefteqn{P(e_1\!+\!e_2\!+\!e_b\!=\!1)-P(e_1\!+\!e_2\!=\!1)} \nonumber \\
&& =-P(e_1\!+\!e_2\!=\!1)\bigl(1-P(e_b\!=\!0)\bigr) \nonumber \\
&& \qquad +P(e_1\!+\!e_2\!=\!0)P(e_b\!=\!1) \nonumber \\
&& =P(e_b\!=\!1)\bigl(P(e_1\!+\!e_2\!=\!0)-P(e_1\!+\!e_2\!=\!1)\bigr) \nonumber \\
&& =P(e_b\!=\!1)(1-2\epsilon)^2 \geq 0~(0 \leq \epsilon \leq 1/2) .
\end{eqnarray}
\end{IEEEproof}
\par
{\it Example 1:} Consider the QLI code $C_1$ defined by $G(D)\!=\!(1\!+\!D\!+\!D^2, 1\!+\!D^2)~(L\!=\!1)$.
From an invariant-factor decomposition of $G(D)$,
\begin{equation}
G^{-1}(D)=\left(
\begin{array}{c}
D \\
1\!+\!D
\end{array}
\right)
\end{equation}
is obtained. Hence, we have
\begin{eqnarray}
F+D^LG^{-1} &=& \left(
\begin{array}{c}
1 \\
1
\end{array}
\right)+D\left(
\begin{array}{c}
D \\
1\!+\!D
\end{array}
\right) \nonumber \\
&=& \left(
\begin{array}{c}
1\!+\!D^2 \\
1\!+\!D\!+\!D^2
\end{array}
\right) \nonumber \\
&=& H^T .
\end{eqnarray}
\par
First compare the two estimates of $i_{k-1}$. Note the following:
\begin{eqnarray}
\hat i(k-1 \vert k) &=& z_k^{(1)h}+z_k^{(2)h}=i_{k-1}+\mbox{\boldmath $e$}_kF \nonumber \\
\hat i(k-1 \vert k-1) &=& z_{k-2}^{(1)h}\!+\!z_{k-2}^{(2)h}\!+\!z_{k-1}^{(2)h}\!=\!i_{k-1}\!+\!\mbox{\boldmath $e$}_{k-1}G^{-1} . \nonumber
\end{eqnarray}
From the first equation, the error probability of $\hat i(k-1 \vert k)$ is given by
\begin{eqnarray}
p_s &\stackrel{\triangle}{=}& P(e_k^{(1)}+e_k^{(2)}=1) \nonumber \\
&=& 2\epsilon-2\epsilon^2 .
\end{eqnarray}
\par
On the other hand, from the second equation, the error probability of $\hat i(k-1 \vert k-1)$ is given by
\begin{eqnarray}
p_f &\stackrel{\triangle}{=}& P(e_{k-2}^{(1)}+e_{k-2}^{(2)}+e_{k-1}^{(2)}=1) \nonumber \\
&=& 3\epsilon-6\epsilon^2+4\epsilon^3 .
\end{eqnarray}
Hence, we have
\begin{eqnarray}
p_f-p_s &=& (3\epsilon-6\epsilon^2+4\epsilon^3)-(2\epsilon-2\epsilon^2) \nonumber \\
&=& \epsilon(1-2\epsilon)^2 \geq 0 .
\end{eqnarray}
This inequality implies that $\hat i(k-1 \vert k)$ has higher accuracy as compared with $\hat i(k-1 \vert k-1)$.
\par
Next, we show an example of encoding (see Table I). In this example, the encoder is terminated in state $(00)$ at $k\!=\!8$. In Table I, ``$^*$'' denotes that the information $i_{k-1}$ and its estimate are different. We observe that the relation
\begin{displaymath}
\hat i(k-1 \vert k)=\hat i(k-1 \vert k-1)+\zeta_k
\end{displaymath}
actually holds.
\begin{table}[tb]
\caption{An example of encoding based on $G(D)\!=\!(1\!+\!D\!+\!D^2, 1\!+\!D^2)$}
\label{Table 1}
\begin{center}
\begin{tabular}{c*{8}{|c}}
$k$ & $1$ & $2$ & $3$ & $4$ & $5$ & $6$ & $7$ & $8$ \\
\hline
$i_k$ & $1$ & $0$ & $0$ & $1$ & $0$ & $1$ & $0$ & $0$ \\
$\mbox{\boldmath $y$}_k$ & $11$ & $10$ & $11$ & $11$ & $10$ & $00$ & $10$ & $11$ \\
$\mbox{\boldmath $e$}_k$ & $00$ & $10$ & $00$ & $01$ & $00$ & $10$ & $00$ & $00$ \\
$\mbox{\boldmath $z$}_k^h$ & $11$ & $00$ & $11$ & $10$ & $10$ & $10$ & $10$ & $11$ \\
$\zeta_k$ &  $0$ & $1$ & $0$ & $0$ & $1$ & $0$ & $0$ & $1$ \\
$\hat i(k-1 \vert k)$ & $0$ & $0^*$ & $0$ & $1^*$ & $1$ & $1^*$ & $1$ & $0$ \\
$\hat i(k-1 \vert k-1)$ & $0$ & $1$ & $0$ & $1^*$ & $0^*$ & $1^*$ & $1$ & $1^*$ \\
$i_{k-1}$ & $0$ & $1$ & $0$ & $0$ & $1$ & $0$ & $1$ & $0$
\end{tabular}
\end{center}
\end{table}


\section{Distributions Related to the Main Decoder in an SST Viterbi Decoder}
It is stated~\cite{kai 681} that the innovations approach to linear filtering problems is first to convert the observed process to a white-noise process, and then to treat the resulting simpler white-noise observations problem. In our case, we think this corresponds to the reduction of decoding complexity in the main decoder in an SST Viterbi decoder. We also think the reduction is caused by biased distributions related to the main decoder. First we show that the distribution of the input to the main decoder is biased under low to moderate channel noise level. Next, we show that the state distribution in the code trellis for the main decoder is also biased under the same channel conditions. In either case, a QLI code is used in the discussion. This is because a QLI code is regarded as a general code as well and then we can compare two distributions, i.e., the one obtained as a general code and the other obtained as a QLI code. Furthermore, we show that the state distribution in the error trellis is equally biased.

\subsection{Information Obtained through Observations~\cite{ari 77}}
Consider the channel model in Section I:
\begin{displaymath}
z_j=x_j\sqrt{2E_s/N_0}+w_j=cx_j+w_j ,
\end{displaymath}
where $c\stackrel{\triangle}{=}\sqrt{2E_s/N_0}$. The conditional entropy $H[z \vert x]$ of the observation $z_j$ given $x_j$ is equal to the entropy $H[w]$ of $w_j$, where $H[w]$ is given by
\begin{equation}
H[w]=\frac{1}{2} \log(2 \pi e) .
\end{equation}
\par
Suppose that $y_j$ has values $0$ and $1$ with equal probability. Then the probability density function of $z_j$, denoted $p(y)$, is given by
\begin{equation}
p(y)=\frac{1}{2}q(y-c)+\frac{1}{2}q(y+c) ,
\end{equation}
where
\begin{displaymath}
q(y)=\frac{1}{\sqrt{2\pi}}e^{-\frac{y^2}{2}} .
\end{displaymath}
\par
{\it Remark 1:} When there is no danger of confusion, we call the {\it probability density function} of a random variable $X$ simply the {\it distribution} of $X$.
\par
Let us calculate the entropy $H[z]$ of $z_j$~\cite{vit 79}. Since
\begin{displaymath}
\int_{-\infty}^{\infty}yq(y)dy=\frac{c}{2}+\frac{(-c)}{2}=0
\end{displaymath}
and
\begin{displaymath}
\int_{-\infty}^{\infty}y^2q(y)dy=\frac{1+c^2}{2}+\frac{1+c^2}{2}=1+c^2 ,
\end{displaymath}
the entropy $H[z]$ associated with $p(y)$~\cite{vit 79} is computed as
\begin{equation}
H[z]=-\int_{-\infty}^{\infty}p(y)\log p(y)dy \leq \frac{1}{2} \log \bigl(2\pi e(1+c^2)\bigr) ,
\end{equation}
with equality when $p(y)$ is Gaussian.
\par
Hence, we have
\begin{eqnarray}
H[x; z] &=& H[z]-H[w] \nonumber \\
&\leq& \frac{1}{2} \log \bigl(2\pi e(1+c^2)\bigr)-\frac{1}{2} \log(2 \pi e) \nonumber \\
&=& \frac{1}{2}\log(1+c^2) ,
\end{eqnarray}
where $H[x; z]$ represents the information obtained through the observation~\cite{ari 77}.
\par
{\it Remark 2:} $H[x; z]$ is the channel capacity of the binary-input AWGN channel~\cite{vit 79}DSuppose that $c \rightarrow 0~(\sqrt{2E_s/N_0} \rightarrow 0)$. Then the inequality almost becomes an equality. Also, note that
\begin{displaymath}
\log(1+c^2) \approx c^2~(c \rightarrow 0) .
\end{displaymath}
Then we have
\begin{equation}
H[x; z] \approx \frac{1}{2}2E_s/N_0=E_s/N_0~(c \rightarrow 0) .
\end{equation}

\subsection{Entropy Associated with the Distribution of the Input to the Main Decoder}
\subsubsection{General codes}
Suppose that the inverse encoder $G^{-1}(D)$ is used as a pre-decoder. Let $\mbox{\boldmath $r$}_k\!=\!(r_k^{(1)}, \cdots, r_k^{(n_0)})$ be the input to the main decoder in an SST Viterbi decoder. We have the following.
\begin{pro}
The distribution of $r_k^{(l)}~(1 \leq l \leq n_0)$ is given by
\begin{equation}
p_r(y)=(1-\alpha)q(y-c)+\alpha q(y+c) ,
\end{equation}
where
\begin{equation}
\alpha\stackrel{\triangle}{=}P(e_k^{(l)}\!=\!0, r_k^{(l)h}\!=\!1)+P(e_k^{(l)}\!=\!1, r_k^{(l)h}\!=\!0) .
\end{equation}
\end{pro}
\begin{IEEEproof}
We can assume that the all-zero code sequence is transmitted. In this case, the distribution of $z_k^{(l)}$ is given by $q(y-c)$ and we have
\begin{displaymath}
z_k^{(l)}=\left\{
\begin{array}{rl}
\vert z_k^{(l)} \vert,& \quad e_k^{(l)}=0 \\
- \vert z_k^{(l)} \vert,& \quad e_k^{(l)}=1 .
\end{array} \right.
\end{displaymath}
On the other hand, from the structure of the SST Viterbi decoder (cf. Fig.1), it follows that
\begin{displaymath}
r_k^{(l)}=\left\{
\begin{array}{rl}
\vert z_k^{(l)} \vert,& \quad r_k^{(l)h}=0 \\
- \vert z_k^{(l)} \vert,& \quad r_k^{(l)h}=1 .
\end{array} \right.
\end{displaymath}
Hence, there are four cases:
\begin{itemize}
\item[1)] $e_k^{(l)}\!=\!0, r_k^{(l)h}\!=\!0 \rightarrow z_k^{(l)}\!=\!\vert z_k^{(l)} \vert, r_k^{(l)}\!=\!\vert z_k^{(l)} \vert$
\item[2)] $e_k^{(l)}\!=\!0, r_k^{(l)h}\!=\!1 \rightarrow z_k^{(l)}\!=\!\vert z_k^{(l)} \vert, r_k^{(l)}\!=\!-\vert z_k^{(l)} \vert$
\item[3)] $e_k^{(l)}\!=\!1, r_k^{(l)h}\!=\!0 \rightarrow z_k^{(l)}\!=\!-\vert z_k^{(l)} \vert, r_k^{(l)}\!=\!\vert z_k^{(l)} \vert$
\item[4)] $e_k^{(l)}\!=\!1, r_k^{(l)h}\!=\!1 \rightarrow z_k^{(l)}\!=\!-\vert z_k^{(l)} \vert, r_k^{(l)}\!=\!-\vert z_k^{(l)} \vert$.
\end{itemize}
In cases 2) and 3), $r_k^{(l)}\!=\!-z_k^{(l)}$ holds and the distribution of $r_k^{(l)}\!=\!-z_k^{(l)}$ becomes $q(y+c)$. Hence, the distribution of $r_k^{(l)}$ is given by
\begin{displaymath}
p_r(y)=(1-\alpha)q(y-c)+\alpha q(y+c) ,
\end{displaymath}
where
\begin{displaymath}
\alpha=P(e_k^{(l)}\!=\!0, r_k^{(l)h}\!=\!1)+P(e_k^{(l)}\!=\!1, r_k^{(l)h}\!=\!0) .
\end{displaymath}
\end{IEEEproof}
\par
Next, let us calculate the entropy of $r_k^{(l)}$, denoted $H[r]$. For the purpose, we calculate the variance $\sigma_r^2$ of $p_r(y)$. Note the following:
\begin{eqnarray}
m_r &=& \int_{-\infty}^{\infty}yp_r(y)dy \nonumber \\
&=& (1-\alpha)\int_{-\infty}^{\infty}yq(y-c)dy+\alpha \int_{-\infty}^{\infty}yq(y+c)dy \nonumber \\
&=& (1-\alpha)c+\alpha (-c) \nonumber \\
&=& c(1-2 \alpha) \nonumber
\end{eqnarray}
\begin{eqnarray}
\lefteqn{\int_{-\infty}^{\infty}y^2p_r(y)dy} \nonumber \\
&& =(1-\alpha)\int_{-\infty}^{\infty}y^2 q(y-c)dy+\alpha \int_{-\infty}^{\infty}y^2 q(y+c)dy \nonumber \\
&& =(1-\alpha)(1+c^2)+\alpha (1+c^2) \nonumber \\
&& =1+c^2 . \nonumber
\end{eqnarray}
Then
\begin{eqnarray}
\sigma_r^2 &=& \int_{-\infty}^{\infty}y^2p_r(y)dy-m_r^2 \nonumber \\
&=& (1+c^2)-c^2(1-2 \alpha)^2 \nonumber \\
&=& 1+4c^2 \alpha(1-\alpha) \nonumber
\end{eqnarray}
is obtained. Hence, we have
\begin{eqnarray}
H[r] &=& -\int_{-\infty}^{\infty}p_r(y)\log p_r(y)dy \nonumber \\
&\leq& \frac{1}{2} \log \bigl(2\pi e(1+4c^2 \alpha(1-\alpha))\bigr) ,
\end{eqnarray}
with equality when $p_r(y)$ is Gaussian. We remark that the right-hand side contains a parameter $\alpha$ which depends on $e_k^{(l)}$ and $r_k^{(l)h}$. Hence, $\alpha$ inevitably depends on $G(D)$ (cf. $\mbox{\boldmath $r$}_k^h\!=\!\mbox{\boldmath $e$}_k(G^{-1}G\!+\!I_{n_0})$).
\par
We have already calculated $H[z]$ and $H[r]$. However, all of the obtained expressions are inequalities. First consider the difference $H[z]-H[r]$. Let $\epsilon\!=\!Q(\sqrt{2E_s/N_0})$ be the channel error probability. We need the following.
\newtheorem{lem}[df]{Lemma}
\begin{lem}
For $0 \leq \epsilon \leq 1/2$, we have $0 \leq \alpha \leq 1/2$.
\end{lem}
\begin{IEEEproof}
See Appendix A.
\end{IEEEproof}
\par
Note that $p_r(y)$ is biased and that the smaller $\epsilon$ becomes (i.e., $\alpha \rightarrow 0$), the more $p_r(y)$ is biased. Hence, it is expected that $H[z]-H[r] \geq 0$ and $H[z]-H[r]$ increases as $\epsilon$ decreases.
\par
On the other hand, let us evaluate the difference between the right-hand sides of $H[z]$ and $H[r]$, i.e.,
\begin{eqnarray}
\lefteqn{\frac{1}{2} \log \bigl(2\pi e(1+c^2)\bigr)} \nonumber \\
&& \qquad -\frac{1}{2} \log \bigl(2\pi e(1+4c^2 \alpha(1-\alpha))\bigr) \nonumber \\
&& =\frac{1}{2}\log \left(\frac{1+c^2}{1+4c^2 \alpha(1-\alpha)}\right) .
\end{eqnarray}
Since $0 \leq \alpha \leq 1/2$, we have
\begin{displaymath}
0 \leq 4 \alpha(1-\alpha) \leq 1 .
\end{displaymath}
Hence, from
\begin{displaymath}
1+4c^2 \alpha(1-\alpha) \leq 1+c^2 ,
\end{displaymath}
it follows that
\begin{displaymath}
\frac{1}{2}\log \frac{1+c^2}{1+4c^2 \alpha(1-\alpha)}=\frac{1}{2}\log (1+\theta)~(\theta \geq 0) .
\end{displaymath}
\par
Moreover, consider the special cases, 1) $\epsilon \rightarrow 0$ and 2) $\epsilon \rightarrow 1/2$. 
\begin{itemize}
\item [1)] $\epsilon \rightarrow 0$: We see that $p_r(y) \rightarrow q(y-c)$, where $q(y-c)$ is Gaussian. Hence, we have 
\begin{displaymath}
H[r]\approx \frac{1}{2} \log \bigl(2\pi e(1+4c^2 \alpha(1-\alpha))\bigr) .
\end{displaymath}
Then we approximately have
\begin{eqnarray}
H[z]-H[r] &\leq& \frac{1}{2}\log \left(\frac{1+c^2}{1+4c^2 \alpha(1-\alpha)}\right) \nonumber \\
&\approx& \frac{1}{2}\log(1+c^2)~(c \rightarrow \infty) . \nonumber
\end{eqnarray}
\item [2)] $\epsilon \rightarrow 1/2$: We see that $p(y) \rightarrow q(y)$, where $q(y)$ is Gaussian. Hence, we have
\begin{displaymath}
H[z]\approx \frac{1}{2} \log \bigl(2\pi e(1+c^2)\bigr) .
\end{displaymath}
Then we approximately have
\begin{eqnarray}
H[z]-H[r] &\geq& \frac{1}{2}\log \left(\frac{1+c^2}{1+4c^2 \alpha(1-\alpha)}\right) \nonumber \\
&\approx& \frac{1}{2}\log \left(\frac{1+c^2}{1+c^2}\right)=0~(c \rightarrow 0) . \nonumber
\end{eqnarray}
\end{itemize}
\par
Furthermore, observe that as $\epsilon~(0 \leq \epsilon \leq 1/2)$ decreases, $\frac{1}{2}\log \left(\frac{1+c^2}{1+4c^2 \alpha(1-\alpha)}\right)$ increases (cf. Table II). We see that this is consistent with the expected behavior of $H[z]-H[r]$.
\par
We have not derived the exact value of $H[z]-H[r]$. However, the above argument implies that the two quantities $H[z]-H[r]$ and $\frac{1}{2}\log \left(\frac{1+c^2}{1+4c^2 \alpha(1-\alpha)}\right)$ have a close relation and the latter can be regarded as an approximation of $H[z]-H[r]$. Hence, in the following, we will compute the latter in order to evaluate $H[z]-H[r]$. Also, the relationship between the two quantities is denoted as
\begin{equation}
H[z]-H[r] \approx \frac{1}{2}\log \left(\frac{1+c^2}{1+4c^2 \alpha(1-\alpha)}\right) ,
\end{equation}
where the notation ``$\approx$'' is used in the above meaning.
\par
We remark that the above calculation applies to a single component of the branch code. However, in order to know the bias of the composite distribution, we should calculate the entropy corresponding to the whole branch. Note that in our channel model, the branch code is transmitted symbol by symbol. Then the distributions corresponding to each code symbol are statistically independent of each other. Hence, the entropy associated with the composite distribution, denoted $H[r_1, r_2, \cdots, r_{n_0}]$, is the sum of the entropies associated with the distributions corresponding to each code symbol. That is, we have
\begin{equation}
H[r_1, r_2, \cdots, r_{n_0}]=H[r_1]\!+\!H[r_2]\!+\!\cdots \!+\!H[r_{n_0}] .
\end{equation}

\subsubsection{QLI codes}
Let
\begin{displaymath}
G(D)=(g_1(D), g_2(D))~(g_1+g_2=D^L)
\end{displaymath}
be a generator matrix for a QLI code. Suppose that $F\!=\!(1, 1)^T$ is used as a pre-decoder. Let $\mbox{\boldmath $\eta$}_{k-L}\!=\!(\eta_{k-L}^{(1)}, \eta_{k-L}^{(2)})$ be the input to the main-decoder in an SST Viterbi decoder (see Fig.2). We have the following.
\begin{pro}
The distribution of $\eta_{k-L}^{(l)}~(l\!=\!1, 2)$ is given by
\begin{equation}
p_{\eta}(y)=(1-\beta)q(y-c)+\beta q(y+c) ,
\end{equation}
where
\begin{equation}
\beta\stackrel{\triangle}{=}P(e_{k-L}^{(l)}\!=\!0, \zeta_k\!=\!1)+P(e_{k-L}^{(l)}\!=\!1, \zeta_k\!=\!0) .
\end{equation}
\end{pro}
\begin{IEEEproof}
Suppose that the all-zero code sequence is transmitted as before. In this case, the distribution of $z_{k-L}^{(l)}$ is given by $q(y-c)$ and we have
\begin{displaymath}
z_{k-L}^{(l)}=\left\{
\begin{array}{rl}
\vert z_{k-L}^{(l)} \vert,& \quad e_{k-L}^{(l)}=0 \\
- \vert z_{k-L}^{(l)} \vert,& \quad e_{k-L}^{(l)}=1 .
\end{array} \right.
\end{displaymath}
On the other hand, we already have $\mbox{\boldmath $\eta$}_{k-L}^h\!=\!(\zeta_k, \zeta_k)$. Then it follows that
\begin{displaymath}
\eta_{k-L}^{(l)}=\left\{
\begin{array}{rl}
\vert z_{k-L}^{(l)} \vert,& \quad \zeta_k=0 \\
- \vert z_{k-L}^{(l)} \vert,& \quad \zeta_k=1 .
\end{array} \right.
\end{displaymath}
Hence, there are four cases:
\begin{itemize}
\item[1)] $e_{k-L}^{(l)}\!=\!0, \zeta_k=0 \rightarrow z_{k-L}^{(l)}\!=\!\vert z_{k-L}^{(l)} \vert, \eta_{k-L}^{(l)}\!=\!\vert z_{k-L}^{(l)} \vert$
\item[2)] $e_{k-L}^{(l)}\!=\!0, \zeta_k=1 \rightarrow z_{k-L}^{(l)}\!=\!\vert z_{k-L}^{(l)} \vert, \eta_{k-L}^{(l)}\!=\!-\vert z_{k-L}^{(l)} \vert$
\item[3)] $e_{k-L}^{(l)}\!=\!1, \zeta_k=0 \rightarrow z_{k-L}^{(l)}\!=\!-\vert z_{k-L}^{(l)} \vert, \eta_{k-L}^{(l)}\!=\!\vert z_{k-L}^{(l)} \vert$
\item[4)] $e_{k-L}^{(l)}\!=\!1, \zeta_k=1 \rightarrow z_{k-L}^{(l)}\!=\!-\vert z_{k-L}^{(l)} \vert, \eta_{k-L}^{(l)}\!=\!-\vert z_{k-L}^{(l)} \vert$.
\end{itemize}
In cases 2) and 3), $\eta_{k-L}^{(l)}\!=\!-z_{k-L}^{(l)}$ holds and the distribution of $\eta_{k-L}^{(l)}\!=\!-z_{k-L}^{(l)}$ becomes $q(y+c)$. Hence, the distribution of $\eta_{k-L}^{(l)}$ is given by
\begin{displaymath}
p_{\eta}(y)=(1-\beta)q(y-c)+\beta q(y+c) ,
\end{displaymath}
where
\begin{displaymath}
\beta=P(e_{k-L}^{(l)}\!=\!0, \zeta_k\!=\!1)+P(e_{k-L}^{(l)}\!=\!1, \zeta_k\!=\!0) .
\end{displaymath}
\end{IEEEproof}
\par
The rest of the argument follows as in the preceding section. Let $H[\eta]$ be the entropy of $\eta_{k-L}^{(l)}$. Then we have
\begin{eqnarray}
H[\eta] &=& -\int_{-\infty}^{\infty}p_{\eta}(y)\log p_{\eta}(y)dy \nonumber \\
&\leq& \frac{1}{2} \log (2\pi e(1+4c^2 \beta(1-\beta))) ,
\end{eqnarray}
with equality when $p_{\eta}(y)$ is Gaussian. Also, we have
\begin{eqnarray}
H[z]-H[\eta] &\approx& \frac{1}{2} \log (2\pi e(1+c^2)) \nonumber \\
&& -\frac{1}{2} \log (2\pi e(1+4c^2 \beta(1-\beta))) \nonumber \\
&=& \frac{1}{2}\log \left(\frac{1+c^2}{1+4c^2 \beta(1-\beta)}\right) ,
\end{eqnarray}
where the notation ``$\approx$'' is employed in the same meaning as in the case of general codes. Furthermore, we have used the following (cf. Lemma 3.2).
\begin{lem}
For $0 \leq \epsilon \leq 1/2$, we have $0 \leq \beta \leq 1/2$.
\end{lem}
\begin{IEEEproof}
See Appendix B.
\end{IEEEproof}

\subsubsection{An example}
Consider the QLI code $C_1$ defined in Example 1. First we regard $C_1$ as a general code ($G^{-1}$ is used as a pre-decoder). Let us evaluate the parameter $\alpha$ defined in the previous section. For the first component of a branch, we have
\begin{displaymath}
\alpha_1=5\epsilon-20\epsilon^2+40\epsilon^3-40\epsilon^4+16\epsilon^5 ,
\end{displaymath}
where $\epsilon\!=\!Q(\sqrt{2E_s/N_0})\!=\!Q(\sqrt{E_b/N_0})$ is the channel error probability. Similarly, for the second component of the branch, we have
\begin{displaymath}
\alpha_2=6\epsilon-30\epsilon^2+80\epsilon^3-120\epsilon^4+96\epsilon^5-32\epsilon^6 .
\end{displaymath}
Hence,
\begin{equation}
H_r^{(1)}\stackrel{\triangle}{=}H[z_1]-H[r_1]\approx \frac{1}{2}\log \left(\frac{1+c^2}{1+4c^2 \alpha_1(1-\alpha_1)}\right)
\end{equation}
\begin{equation}
H_r^{(2)}\stackrel{\triangle}{=}H[z_2]-H[r_2]\approx \frac{1}{2}\log \left(\frac{1+c^2}{1+4c^2 \alpha_2(1-\alpha_2)}\right)
\end{equation}
are obtained, where $c\!=\!\sqrt{2E_s/N_0}\!=\!\sqrt{E_b/N_0}$.
\par
Next, we regard $C_1$ as a QLI code ($F$ is used as a pre-decoder) and evaluate the parameter $\beta$. In this case, we have
\begin{displaymath}
\beta_1=6\epsilon-30\epsilon^2+80\epsilon^3-120\epsilon^4+96\epsilon^5-32\epsilon^6~~(=\alpha_2)
\end{displaymath}
for the first component of a branch. Similarly, for the second component of the branch, we have
\begin{displaymath}
\beta_2=4\epsilon-12\epsilon^2+16\epsilon^3-8\epsilon^4 .
\end{displaymath}
Hence,
\begin{equation}
H_{\eta}^{(1)}\stackrel{\triangle}{=}H[z_1]-H[\eta_1]\approx \frac{1}{2}\log \left(\frac{1+c^2}{1+4c^2 \beta_1(1-\beta_1)}\right)
\end{equation}
\begin{equation}
H_{\eta}^{(2)}\stackrel{\triangle}{=}H[z_2]-H[\eta_2]\approx \frac{1}{2}\log \left(\frac{1+c^2}{1+4c^2 \beta_2(1-\beta_2)}\right)
\end{equation}
are obtained.
\par
Tables II and III show entropy versus $E_b/N_0$. From these tables, we observe that
\begin{equation}
H_r^{(1)}+H_r^{(2)} < H_{\eta}^{(1)}+H_{\eta}^{(2)} .
\end{equation}
That is, when $C_1$ is regarded as a QLI code, the distribution of the input to the main decoder is more biased.
\begin{table}[tb]
\caption{Entropies associated with input distributions (as a general code)}
\label{Table 2}
\begin{center}
\scalebox{0.8}{
\begin{tabular}{c*{7}{|c}}
$E_b/N_0\,(\mbox{dB})$ & $c$ & $\epsilon$ & $\alpha_1$ & $\alpha_2$ & $H_r^{(1)}$ & $H_r^{(2)}$ & $H_r^{(1)}\!+\!H_r^{(2)}$ \\
\hline
$0$ & $1.000$ & $0.1587$ & $0.4259$ & $0.4494$ & $0.0055$ & $0.0026$ & $0.0081$ \\
$1$ & $1.122$ & $0.1309$ & $0.3904$ & $0.4191$ & $0.0136$ & $0.0075$ & $0.0211$ \\
$2$ & $1.259$ & $0.1040$ & $0.3442$ & $0.3766$ & $0.0307$ & $0.0190$ & $0.0497$ \\
$3$ & $1.413$ & $0.0788$ & $0.2879$ & $0.3213$ & $0.0639$ & $0.0445$ & $0.1084$ \\
$4$ & $1.585$ & $0.0565$ & $0.2255$ & $0.2565$ & $0.1214$ & $0.0929$ & $0.2143$ \\
$5$ & $1.778$ & $0.0377$ & $0.1621$ & $0.1876$ & $0.2131$ & $0.1759$ & $0.3890$ \\
$6$ & $1.995$ & $0.0230$ & $0.1049$ & $0.1231$ & $0.3456$ & $0.3027$ & $0.6483$ \\
$7$ & $2.239$ & $0.0126$ & $0.0599$ & $0.0710$ & $0.5191$ & $0.4756$ & $0.9947$ \\
$8$ & $2.512$ & $0.00600$ & $0.0293$ & $0.0349$ & $0.7241$ & $0.6870$ & $1.4111$ \\
$9$ & $2.818$ & $0.00242$ & $0.0120$ & $0.0143$ & $0.9355$ & $0.9103$ & $1.8458$ \\
$10$ & $3.162$ & $0.00078$ & $0.0039$ & $0.0047$ & $1.1266$ & $1.1131$ & $2.2397$
\end{tabular}
}
\end{center}
\end{table}
\begin{table}[tb]
\caption{Entropies associated with input distributions (as a QLI code)}
\label{Table 3}
\begin{center}
\scalebox{0.8}{
\begin{tabular}{c*{7}{|c}}
$E_b/N_0\,(\mbox{dB})$ & $c$ & $\epsilon$ & $\beta_1$ & $\beta_2$ & $H_{\eta}^{(1)}$ & $H_{\eta}^{(2)}$ & $H_{\eta}^{(1)}\!+\!H_{\eta}^{(2)}$ \\
\hline
$0$ & $1.000$ & $0.1587$ & $0.4494$ & $0.3914$ & $0.0026$ & $0.0119$ & $0.0145$ \\
$1$ & $1.122$ & $0.1309$ & $0.4191$ & $0.3515$ & $0.0075$ & $0.0252$ & $0.0327$ \\
$2$ & $1.259$ & $0.1040$ & $0.3766$ & $0.3033$ & $0.0190$ & $0.0498$ & $0.0688$ \\
$3$ & $1.413$ & $0.0788$ & $0.3213$ & $0.2482$ & $0.0445$ & $0.0926$ & $0.1371$ \\
$4$ & $1.585$ & $0.0565$ & $0.2565$ & $0.1905$ & $0.0929$ & $0.1602$ & $0.2531$ \\
$5$ & $1.778$ & $0.0377$ & $0.1876$ & $0.1346$ & $0.1759$ & $0.2602$ & $0.4361$ \\
$6$ & $1.995$ & $0.0230$ & $0.1231$ & $0.0858$ & $0.3027$ & $0.3975$ & $0.7002$ \\
$7$ & $2.239$ & $0.0126$ & $0.0710$ & $0.0485$ & $0.4756$ & $0.5694$ & $1.0450$ \\
$8$ & $2.512$ & $0.00600$ & $0.0349$ & $0.0236$ & $0.6870$ & $0.7654$ & $1.4524$ \\
$9$ & $2.818$ & $0.00242$ & $0.0143$ & $0.0096$ & $0.9103$ & $0.9634$ & $1.8737$ \\
$10$ & $3.162$ & $0.00078$ & $0.0047$ & $0.0031$ & $1.1131$ & $1.1406$ & $2.2536$
\end{tabular}
}
\end{center}
\end{table}

\subsection{State Distribution in the Code Trellis for the Main Decoder}
In the preceding section, it was shown that the distribution of the input to the main decoder is biased under moderately noisy conditions. In this section, we show that the state distribution in the code trellis for the main decoder is also biased under the same channel conditions. For the purpose, we will take a QLI code. Since a QLI code can be regarded as a general code as well, we have two state expressions for the main decoder. Hence, we can evaluate a likelihood concentration in the main decoder more precisely by comparing the two state distributions.
\par
{\it Remark 1:} Note that the code trellis module can be constructed as an error trellis module based on the syndrome former. We remark that for a high-rate code, the resulting code trellis module has less complexity than that of the conventional one~\cite{sido 94,yamada 83}. Lee et al.~\cite{lee 96} used this method when they applied the SST scheme to $(n_0, n_0-1)$ convolutional codes.
\par
Consider a QLI code defined by $G(D)\!=\!(g_1(D), g_2(D))$. A likelihood concentration in the main decoder depends heavily on the choice of a pre-decoder. Roughly speaking, if the information $u_k$ for the main decoder consists of smaller number of error terms, then a higher likelihood concentration occurs. First apply $F$ as a pre-decoder. Then we have
\begin{equation}
u_k=e_k^{(1)}+e_k^{(2)}
\end{equation}
and $u_k$ consists of two error terms. Next, apply the inverse encoder $G^{-1}$ as a pre-decoder. Suppose that
\begin{equation}
G^{-1}=\left(
\begin{array}{c}
b_1 \\
b_2
\end{array}
\right) ,
\end{equation}
where $b_1$ and $b_2$ are polynomials in $D$. If these polynomials consist of small number of terms, then $u_k\!=\!\mbox{\boldmath $e$}_kG^{-1}$ also consists of small number of error terms, which results in a high likelihood concentration in the main decoder. Let $n_e$ be the number of error terms in $u_k$. Since $n_e>2$ in general, QLI codes are preferable from a likelihood concentration viewpoint. On the other hand, for any fixed $\nu$, the free distance, denoted $d_{free}$, of the best QLI codes is a little less than that of the best overall codes. (Here the optimality criterion first maximizes $d_{free}$ and then minimizes $N_{d_{free}}$, where $N_{d_{free}}$ is the number of codewords with weight $d_{free}$~\cite{lin 04}.) In order to cope with this problem in application of the SST scheme, Ping et al.~\cite{pin 91} searched for a good non-systematic encoder whose inverse consists of polynomials with small number of terms. For $\nu\!=\!6$, they found the generator matrix
\begin{equation}
G(D)\!=\!(1\!+\!D\!+\!D^4\!+\!D^5\!+\!D^6, 1\!+\!D^2\!+\!D^3\!+\!D^4\!+\!D^6)
\end{equation}
with
\begin{displaymath}
G^{-1}=\left(
\begin{array}{c}
D \\
1\!+\!D
\end{array}
\right) .
\end{displaymath}
Note that the above $G(D)$ is an {\it optimum distance profile} (ODP) encoding matrix~\cite[Table 8.1]{joha 99} and the corresponding code has $d_{free}\!=\!10$. It is shown that
\begin{equation}
G(D)=(1\!+\!D\!+\!D^4, 1\!+\!D^2\!+\!D^3\!+\!D^4)
\end{equation}
has the same inverse encoder. Note that the above is also an ODP encoding matrix.
\par
{\it Example 2:} Consider the QLI code $C_1$ defined in Example 1. First we regard $C_1$ as a general code ($G^{-1}$ is used as a pre-decoder). In this case, the information $u_k$ for the main decoder is given by
\begin{eqnarray}
u_k &=& \mbox{\boldmath $e$}_k\left(
\begin{array}{c}
D \\
1\!+\!D
\end{array}
\right) \nonumber \\
&=& e_{k-1}^{(1)}+e_{k-1}^{(2)}+e_k^{(2)} .
\end{eqnarray}
Accordingly, the trellis state becomes
\begin{displaymath}
\mbox{\boldmath $s$}_k\!=\!(u_{k-1}, u_k)\!=\!(e_{k-2}^{(1)}\!+\!e_{k-2}^{(2)}\!+\!e_{k-1}^{(2)}, e_{k-1}^{(1)}\!+\!e_{k-1}^{(2)}\!+\!e_k^{(2)}) .
\end{displaymath}
Hence, we have
\begin{eqnarray}
P_{00}\stackrel{\triangle}{=}P(\mbox{\boldmath $s$}_k\!=\!(00)) &=& 1-5\epsilon+12\epsilon^2-12\epsilon^3+4\epsilon^4 \nonumber \\
P_{01}\stackrel{\triangle}{=}P(\mbox{\boldmath $s$}_k\!=\!(01)) &=& 2\epsilon-6\epsilon^2+8\epsilon^3-4\epsilon^4 \nonumber \\
P_{10}\stackrel{\triangle}{=}P(\mbox{\boldmath $s$}_k\!=\!(10)) &=& 2\epsilon-6\epsilon^2+8\epsilon^3-4\epsilon^4 \nonumber \\
P_{11}\stackrel{\triangle}{=}P(\mbox{\boldmath $s$}_k\!=\!(11)) &=& \epsilon-4\epsilon^3+4\epsilon^4 , \nonumber
\end{eqnarray}
where $\epsilon\!=\!Q(\sqrt{E_b/N_0})$ is the channel error probability.
\par
Next, we regard $C_1$ as a QLI code ($F$ is used as a pre-decoder). Then the information $u_k$ for the main decoder is given by
\begin{eqnarray}
u_k &=& \mbox{\boldmath $e$}_k\left(
\begin{array}{c}
1 \\
1
\end{array}
\right) \nonumber \\
&=& e_k^{(1)}+e_k^{(2)} . \nonumber
\end{eqnarray}
Accordingly, the trellis state becomes
\begin{displaymath}
\mbox{\boldmath $s$}_k=(u_{k-1}, u_k)=(e_{k-1}^{(1)}\!+\!e_{k-1}^{(2)}, e_k^{(1)}\!+\!e_k^{(2)}) .
\end{displaymath}
Hence, we have
\begin{eqnarray}
P_{00} &=& 1-4\epsilon+8\epsilon^2-8\epsilon^3+4\epsilon^4 \nonumber \\
P_{01} &=& 2\epsilon-6\epsilon^2+8\epsilon^3-4\epsilon^4 \nonumber \\
P_{10} &=& 2\epsilon-6\epsilon^2+8\epsilon^3-4\epsilon^4 \nonumber \\
P_{11} &=& 4\epsilon^2-8\epsilon^3+4\epsilon^4 . \nonumber
\end{eqnarray}
\par
In either case, the entropy $H$ associated with the state distribution is given by
\begin{eqnarray}
H &=& -P_{00}\log_2P_{00}-P_{01}\log_2P_{01} \nonumber \\
&& \qquad -P_{10}\log_2P_{10}-P_{11}\log_2P_{11} .
\end{eqnarray}
The results are shown in Tables IV and V. We observe that a higher likelihood concentration to state $(00)$ occurs when the code is regarded as a QLI code. Denote by $\mbox{\boldmath $s$}_k^p$ and $\mbox{\boldmath $s$}_k^q$ the states for the main decoder obtained as a general code and as a QLI code, respectively. Note that $u_k$ consists of three error terms in $\mbox{\boldmath $s$}_k^p$, whereas $u_k$ consists of two error terms in $\mbox{\boldmath $s$}_k^q$. As was stated above, a likelihood concentration in the main decoder depends on the number of error terms ($n_e$) forming $u_k$ in general. Hence, the results are reasonable.
\par
{\it Remark 2:} Note that the components of the state are not statistically independent of each other in general. For example, take $\mbox{\boldmath $s$}_k^p\!=\!(u_{k-1}, u_k)\!=\!(e_{k-2}^{(1)}\!+\!e_{k-2}^{(2)}\!+\!e_{k-1}^{(2)}, e_{k-1}^{(1)}\!+\!e_{k-1}^{(2)}\!+\!e_k^{(2)})$. We see that $e_{k-1}^{(2)}$ is contained in both components. Hence, $n_e$ alone does not affect the state distribution. Nevertheless, $n_e$ provides useful information about a likelihood concentration in the main decoder.
\begin{table}[tb]
\caption{State distributions for the main decoder (as a general code)}
\label{Table 4}
\begin{center}
\scalebox{0.9}{
\begin{tabular}{c*{6}{|c}}
$E_b/N_0\,(\mbox{dB})$ & $\epsilon$ & $P_{00}$ & $P_{01}$ & $P_{10}$ & $P_{11}$ & $H$ \\
\hline
$0$ & $0.1587$ & $0.4633$ & $0.1957$ & $0.1957$ & $0.1452$ & $1.8398$ \\
$1$ & $0.1309$ & $0.5253$ & $0.1758$ & $0.1758$ & $0.1231$ & $1.7418$ \\
$2$ & $0.1040$ & $0.5968$ & $0.1516$ & $0.1516$ & $0.1000$ & $1.6019$ \\
$3$ & $0.0788$ & $0.6746$ & $0.1243$ & $0.1243$ & $0.0768$ & $1.4153$ \\
$4$ & $0.0565$ & $0.7536$ & $0.0953$ & $0.0953$ & $0.0558$ & $1.1864$ \\
$5$ & $0.0377$ & $0.8279$ & $0.0673$ & $0.0673$ & $0.0375$ & $0.9273$ \\
$6$ & $0.0230$ & $0.8912$ & $0.0429$ & $0.0429$ & $0.0230$ & $0.6631$ \\
$7$ & $0.0126$ & $0.9389$ & $0.0243$ & $0.0243$ & $0.0126$ & $0.4255$ \\
$8$ & $0.00600$ & $0.9704$ & $0.0118$ & $0.0118$ & $0.0060$ & $0.2376$ \\
$9$ & $0.00242$ & $0.9880$ & $0.0048$ & $0.0048$ & $0.0024$ & $0.1121$ \\
$10$ & $0.00078$ & $0.9961$ & $0.0016$ & $0.0016$ & $0.0008$ & $0.0436$
\end{tabular}
}
\end{center}
\end{table}
\begin{table}[tb]
\caption{State distributions for the main decoder (as a QLI code)}
\label{Table 5}
\begin{center}
\scalebox{0.9}{
\begin{tabular}{c*{6}{|c}}
$E_b/N_0\,(\mbox{dB})$ & $\epsilon$ & $P_{00}$ & $P_{01}$ & $P_{10}$ & $P_{11}$ & $H$ \\
\hline
$0$ & $0.1587$ & $0.5372$ & $0.1957$ & $0.1957$ & $0.0713$ & $1.6745$ \\
$1$ & $0.1309$ & $0.5967$ & $0.1758$ & $0.1758$ & $0.0518$ & $1.5476$ \\
$2$ & $0.1040$ & $0.6620$ & $0.1516$ & $0.1516$ & $0.0347$ & $1.3875$ \\
$3$ & $0.0788$ & $0.7306$ & $0.1243$ & $0.1243$ & $0.0209$ & $1.1953$ \\
$4$ & $0.0565$ & $0.7981$ & $0.0953$ & $0.0953$ & $0.0113$ & $0.9790$ \\
$5$ & $0.0377$ & $0.8601$ & $0.0673$ & $0.0673$ & $0.0053$ & $0.7511$ \\
$6$ & $0.0230$ & $0.9121$ & $0.0429$ & $0.0429$ & $0.0020$ & $0.5288$ \\
$7$ & $0.0126$ & $0.9509$ & $0.0243$ & $0.0243$ & $0.00062$ & $0.3363$ \\
$8$ & $0.00600$ & $0.9763$ & $0.0118$ & $0.0118$ & $0.00014$ & $0.1868$ \\
$9$ & $0.00242$ & $0.9904$ & $0.0048$ & $0.0048$ & $0.000023$ & $0.0882$ \\
$10$ & $0.00078$ & $0.9969$ & $0.0016$ & $0.0016$ & $0.000003$ & $0.0344$
\end{tabular}
}
\end{center}
\end{table}

\subsection{State Distribution in the Error Trellis}
It has been shown~\cite{taji 031,taji 032} that SST Viterbi decoding based on a code trellis and syndrome decoding based on the corresponding error trellis are equivalent. In the following, $k_0$ is assumed to be $(n_0-1)$ for simplicity. Then the size of $H(D)$ is $1 \times n_0$. Let $\nu$ be the constraint length of $H(D)$. Denote by $\mbox{\boldmath $s$}_k$ and $\mbox{\boldmath $\sigma$}_k$ the state at $t\!=\!k$ in the code trellis for the main decoder and the state at $t\!=\!k$ in the error trellis, respectively. Based on an adjoint-obvious realization (observer canonical form~\cite{forn 73}) of the syndrome former $H^T$, $\mbox{\boldmath $\sigma$}_k$ can be expressed as
\begin{equation}
\mbox{\boldmath $\sigma$}_k=\mbox{\boldmath $e$}_kU(D) ,
\end{equation}
where $U(D)$ is an $n_0 \times \nu$ matrix whose entries are polynomials in $D$. Then we have
\begin{eqnarray}
\mbox{\boldmath $\sigma$}_k &=& (\mbox{\boldmath $u$}_kG+\mbox{\boldmath $r$}_k^h)U \nonumber \\
&=& \mbox{\boldmath $u$}_kGU+\zeta_k(H^{-1})^TU .
\end{eqnarray}
Note that the first term $\mbox{\boldmath $u$}_kGU$ corresponds to the syndrome former state obtained by inputting the encoder output $\mbox{\boldmath $u$}_kG$ directly to the syndrome former $H^T$. That is, $\mbox{\boldmath $u$}_kGU$ is the {\it dual (physical) state}~\cite{forn 73} corresponding to the encoder state $\mbox{\boldmath $s$}_k$. Since the space of encoder states and that of the corresponding dual states are isomorphic, the correspondence between $\mbox{\boldmath $s$}_k$ and $\mbox{\boldmath $u$}_kGU$ is one-to-one. Here note that the term $\zeta_k(H^{-1})^TU$ is common to every state $\mbox{\boldmath $s$}_k$. Hence, the correspondence between $\mbox{\boldmath $s$}_k$ and $\mbox{\boldmath $\sigma$}_k$ is also one-to-one. This fact implies that the state distribution in a code trellis for the main decoder is closely related to that in the corresponding error trellis.
\par
{\it Example 2 (Continued):} Consider the QLI code $C_1$ again. Based on an adjoint-obvious realization of the syndrome former $H^T\!=\!\left(
\begin{array}{c}
1\!+\!D^2 \\
1\!+\!D\!+\!D^2
\end{array}
\right)$, the state in the error trellis becomes
\begin{eqnarray}
\mbox{\boldmath $\sigma$}_k &=& (\sigma_{k1}, \sigma_{k2})=(e_{k-1}^{(1)}\!+\!e_{k-1}^{(2)}\!+\!e_k^{(2)}, e_k^{(1)}\!+\!e_k^{(2)}) \nonumber \\
&=& (e_k^{(1)}, e_k^{(2)})\left(
\begin{array}{cc}
D & 1 \\
1\!+\!D & 1
\end{array}
\right) \nonumber \\
&\stackrel{\triangle}{=}& \mbox{\boldmath $e$}_kU(D) .
\end{eqnarray}
Hence, we have
\begin{eqnarray}
\tilde P_{00}\stackrel{\triangle}{=}P(\mbox{\boldmath $\sigma$}_k\!=\!(00)) &=& 1-4\epsilon+7\epsilon^2-4\epsilon^3 \nonumber \\
\tilde P_{01}\stackrel{\triangle}{=}P(\mbox{\boldmath $\sigma$}_k\!=\!(01)) &=& \epsilon-\epsilon^2 \nonumber \\
\tilde P_{10}\stackrel{\triangle}{=}P(\mbox{\boldmath $\sigma$}_k\!=\!(10)) &=& 2\epsilon-5\epsilon^2+4\epsilon^3 \nonumber \\
\tilde P_{11}\stackrel{\triangle}{=}P(\mbox{\boldmath $\sigma$}_k\!=\!(11)) &=& \epsilon-\epsilon^2 , \nonumber
\end{eqnarray}
where $\epsilon\!=\!Q(\sqrt{E_b/N_0})$. The entropy $\tilde H$ associated with the above distribution is given by
\begin{eqnarray}
\tilde H &=& -\tilde P_{00}\log_2 \tilde P_{00}-\tilde P_{01}\log_2 \tilde P_{01} \nonumber \\
&& \qquad -\tilde P_{10}\log_2 \tilde P_{10}-\tilde P_{11}\log_2 \tilde P_{11} .
\end{eqnarray}
The result is shown in Table VI. From Tables IV, V, and VI, we see that $\tilde H$ lies between the value of entropy obtained by regarding $C_1$ as a general code and that obtained by regarding $C_1$ as a QLI code. This observation comes from the state expressions for $\mbox{\boldmath $s$}_k^p$, $\mbox{\boldmath $s$}_k^q$, and $\mbox{\boldmath $\sigma$}_k$:
\begin{eqnarray}
\mbox{\boldmath $s$}_k^p &=&(u_{k-1}, u_k) \nonumber \\
&=& (e_{k-2}^{(1)}\!+\!e_{k-2}^{(2)}\!+\!e_{k-1}^{(2)}, e_{k-1}^{(1)}\!+\!e_{k-1}^{(2)}\!+\!e_k^{(2)}) \nonumber \\
\mbox{\boldmath $s$}_k^q &=& (u_{k-1}, u_k)=(e_{k-1}^{(1)}\!+\!e_{k-1}^{(2)}, e_k^{(1)}\!+\!e_k^{(2)}) \nonumber \\
\mbox{\boldmath $\sigma$}_k &=& (\sigma_{k1}, \sigma_{k2})=(e_{k-1}^{(1)}\!+\!e_{k-1}^{(2)}\!+\!e_k^{(2)}, e_k^{(1)}\!+\!e_k^{(2)}) . \nonumber
\end{eqnarray}
(Also, see Remark 2 at the end of Section III-C.)
\begin{table}[tb]
\caption{State distributions in the error trellis}
\label{Table 6}
\begin{center}
\scalebox{0.9}{
\begin{tabular}{c*{6}{|c}}
$E_b/N_0\,(\mbox{dB})$ & $\epsilon$ & $\tilde P_{00}$ & $\tilde P_{01}$ & $\tilde P_{10}$ & $\tilde P_{11}$ & $\tilde H$ \\
\hline
$0$ & $0.1587$ & $0.5255$ & $0.1335$ & $0.2075$ & $0.1335$ & $1.7344$ \\
$1$ & $0.1309$ & $0.5874$ & $0.1138$ & $0.1851$ & $0.1138$ & $1.6150$ \\
$2$ & $0.1040$ & $0.6552$ & $0.0932$ & $0.1584$ & $0.0932$ & $1.4590$ \\
$3$ & $0.0788$ & $0.7263$ & $0.0726$ & $0.1285$ & $0.0726$ & $1.2649$ \\
$4$ & $0.0565$ & $0.7956$ & $0.0533$ & $0.0978$ & $0.0533$ & $1.0416$ \\
$5$ & $0.0377$ & $0.8589$ & $0.0363$ & $0.0685$ & $0.0363$ & $0.8009$ \\
$6$ & $0.0230$ & $0.9117$ & $0.0225$ & $0.0434$ & $0.0225$ & $0.5645$ \\
$7$ & $0.0126$ & $0.9507$ & $0.0124$ & $0.0244$ & $0.0124$ & $0.3570$ \\
$8$ & $0.00600$ & $0.9763$ & $0.0060$ & $0.0118$ & $0.0060$ & $0.1980$ \\
$9$ & $0.00242$ & $0.9904$ & $0.0024$ & $0.0048$ & $0.0024$ & $0.0926$ \\
$10$ & $0.00078$ & $0.9969$ & $0.0008$ & $0.0016$ & $0.0008$ & $0.0358$
\end{tabular}
}
\end{center}
\end{table}
\par
Finally, examine the correspondence between the state in the code trellis for the main decoder and that in the error trellis. First consider the correspondence between $\mbox{\boldmath $s$}_k^p\!=\!(u_{k-1}, u_k)$ and $\mbox{\boldmath $\sigma$}_k$. Note the relation
\begin{eqnarray}
\mbox{\boldmath $\sigma$}_k &=& (u_kG+\mbox{\boldmath $r$}_k^h)U \nonumber \\
&=& u_kGU+\zeta_k(H^{-1})^TU . \nonumber
\end{eqnarray}
Since
\begin{eqnarray}
GU &=& (1\!+\!D\!+\!D^2, 1\!+\!D^2)\left(
\begin{array}{cc}
D & 1 \\
1\!+\!D & 1
\end{array}
\right) \nonumber \\
&=& (1, D) \nonumber 
\end{eqnarray}
\begin{eqnarray}
(H^{-1})^TU &=& (1\!+\!D, D)\left(
\begin{array}{cc}
D & 1 \\
1\!+\!D & 1
\end{array}
\right) \nonumber \\
&=& (0, 1) , \nonumber
\end{eqnarray}
it follows that
\begin{eqnarray}
\mbox{\boldmath $\sigma$}_k &=& u_kGU+\zeta_k(H^{-1})^TU \nonumber \\
&=& u_k(1, D)+\zeta_k(0, 1) \nonumber \\
&=& (u_k, u_{k-1}\!+\!\zeta_k) . \nonumber
\end{eqnarray}
Hence, we have
\begin{equation}
\mbox{\boldmath $s$}_k^p=(u_{k-1}, u_k)\leftrightarrow\mbox{\boldmath $\sigma$}_k=(u_k, u_{k-1}\!+\!\zeta_k) ,
\end{equation}
where
\begin{eqnarray}
u_{k-1}+\zeta_k &=& (e_{k-2}^{(1)}\!+\!e_{k-2}^{(2)}\!+\!e_{k-1}^{(2)}) \nonumber \\
&& +(e_{k-2}^{(1)}\!+\!e_k^{(1)}\!+\!e_{k-2}^{(2)}\!+\!e_{k-1}^{(2)}\!+\!e_k^{(2)}) \nonumber \\
&=& e_k^{(1)}+e_k^{(2)} . \nonumber
\end{eqnarray}
\par
Next, consider the correspondence between $\mbox{\boldmath $s$}_k^q\!=\!(u_{k-1}, u_k)$ and $\mbox{\boldmath $\sigma$}_k$. This time (cf. Section II-B), note the relation
\begin{eqnarray}
\mbox{\boldmath $\sigma$}_k &=& (u_{k+L}G+\mbox{\boldmath $\eta$}_k^h)U \nonumber \\
&=& u_{k+L}GU+(\zeta_{k+L}, \zeta_{k+L})U . \nonumber
\end{eqnarray}
Letting $L\!=\!1$, it follows that
\begin{eqnarray}
\mbox{\boldmath $\sigma$}_k &=& u_{k+1}GU+(\zeta_{k+1}, \zeta_{k+1})U \nonumber \\
&=& u_{k+1}(1, D)+(\zeta_{k+1}, \zeta_{k+1})\left(
\begin{array}{cc}
D & 1 \\
1\!+\!D & 1
\end{array}
\right) \nonumber \\
&=& (u_{k+1}\!+\!\zeta_{k+1}, u_k) . \nonumber
\end{eqnarray}
Hence, we have
\begin{equation}
\mbox{\boldmath $s$}_k^q=(u_{k-1}, u_k)\leftrightarrow\mbox{\boldmath $\sigma$}_k=(u_{k+1}\!+\!\zeta_{k+1}, u_k) ,
\end{equation}
where
\begin{eqnarray}
u_{k+1}+\zeta_{k+1} &=& (e_{k+1}^{(1)}\!+\!e_{k+1}^{(2)}) \nonumber \\
&& +(e_{k-1}^{(1)}\!+\!e_{k+1}^{(1)}\!+\!e_{k-1}^{(2)}\!+\!e_k^{(2)}\!+\!e_{k+1}^{(2)}) \nonumber \\
&=& e_{k-1}^{(1)}+e_{k-1}^{(2)}+e_k^{(2)} . \nonumber
\end{eqnarray}
These results are consistent with the concrete state expressions for $\mbox{\boldmath $s$}_k^p$, $\mbox{\boldmath $s$}_k^q$, and $\mbox{\boldmath $\sigma$}_k$.


\section{Complexity Reduction in the Main Decoder in an SST Viterbi Decoder}
We have shown that the state distribution in the code trellis for the main decoder in an SST Viterbi decoder is biased under moderately noisy conditions. In this section, we show that those biased distributions actually lead to complexity reduction in the main decoder. Two reduction methods will be discussed. In the first one, biased state distributions are directly used for complexity reduction, whereas in the second one, those distributions are indirectly used. There have been several related works~\cite{arie 95,arie 99,pin 91,taji 86,taji 971,taji 972} since the SST scheme was proposed. Hence, the discussion in the former part is mainly based on these known works. The known material is also dealt with in the latter part, but some original results are contained. In particular, we give an approximate criterion for complexity reduction in the main decoder in relation to the second reduction method.

\subsection{Complexity Reduction Using State Distributions}
So far biased state distributions have been directly used in order to reduce the decoder complexity~\cite{pin 91,taji 86}. In the following, $k_0\!=\!1$ is assumed for simplicity. First we briefly review the {\it generalized Viterbi algorithm} (GVA)~\cite{hashi 87}. Let
\begin{equation}
\mbox{\boldmath $u$}^k\stackrel{\triangle}{=}u_1u_2\cdots u_k
\end{equation}
be the transmitted information sequence, where $k$ is the current depth. In the usual Viterbi algorithm, a trellis diagram is drawn by regarding the latest $\nu$ symbols $(u_{k-\nu+1} \cdots u_k)$ as a state (i.e., encoder state). On the other hand, in the GVA, the latest $\tilde \nu$ symbols $(u_{k-\tilde \nu+1} \cdots u_k)$ is considered as an algorithm's state (i.e., decoder state), where $\tilde \nu~(>0)$ can be chosen independent of $\nu$. $\tilde \nu$ is called a constraint length of the algorithm. By choosing $\tilde \nu$ smaller that $\nu$, the number of decoder states can be reduced. In this case, however, it is not guaranteed that the overall ML path can be chosen if a single survivor is preserved for each decoder state. Note that a decoder state consists of multiple encoder states. Hence, when a survivor for the decoder state is determined, the most likely path for each component encoder state has to be selected beforehand. This procedure is called {\it pre-selection}~\cite{hashi 87}.
\par
In~\cite{taji 86}, the GVA was applied to the main decoder by taking account of a biased state distribution. The method is based on the conjecture that, if a likelihood concentration to some particular states is occurring in the main decoder, then a great deal of decoding complexity reduction can be realized by applying the GVA to the main decoder with $\tilde \nu$ smaller than $\nu$ and by slightly increasing the number of total survivors as compared with that of decoder states. The method is formulated as follows:
\begin{itemize}
\item[1)] The SST scheme is used to produce a likelihood concentration in the main decoder.
\item[2)] The GVA is applied to the main decoder with $\tilde \nu$ smaller than $\nu$.
\item[3)] In order to avoid a performance degradation due to choosing $\tilde \nu$ smaller than $\nu$, more than one survivors are preserved for those decoder states with high probabilities.
\end{itemize}
\par
The above method was applied to the QLI code $C_2$ defined by
\begin{equation}
G(D)\!=\!(1\!+\!D\!+\!D^3\!+\!D^4\!+\!D^6, 1\!+\!D\!+\!D^2\!+\!D^3\!+\!D^4\!+\!D^6) .
\end{equation}
Note that this code has $d_{free}\!=\!9$. We observe that there occurs a likelihood concentration to the all-zero state and the states containing only one ``$1$'' (e.g., $(000001)$). Then $\tilde \nu$ is set to 5 and two survivors are preserved for each of the decoder states with high probabilities and only one survivor for each of the other decoder states. Hence, the number of decoder states is $32$ and $38$ survivors are preserved. Simulation results show that the method can reduce the decoding complexity to almost $1/2$ of that of the conventional one within a very small performance degradation, where $8$-level receiver quantization is assumed. It is also shown that a small increase of the number of survivors (i.e., additional $6$ survivors) significantly improves the performance. This fact comes from a much biased state distribution in the code trellis for the main decoder.
\par
Ping et al.~\cite{pin 91} also used the SST scheme to reduce the decoder complexity. Note that $C_2$ is a QLI code and has not the best $d_{free}$ with $\nu\!=\!6$. On the other hand, the number of error terms in $u_k\!=\!\mbox{\boldmath $e$}_kG^{-1}$ must be small in order to produce a high likelihood concentration in the main decoder. As a result (see Section III-C), they chose the generator matrix
\begin{displaymath}
G(D)\!=\!(1\!+\!D\!+\!D^4\!+\!D^5\!+\!D^6, 1\!+\!D^2\!+\!D^3\!+\!D^4\!+\!D^6)
\end{displaymath}
with
\begin{displaymath}
G^{-1}=\left(
\begin{array}{c}
D \\
1\!+\!D
\end{array}
\right) .
\end{displaymath}
Note that the corresponding code $C_3$ has $d_{free}\!=\!10$. Next, they applied a simplifying scheme to the main decoder. Since the state distribution in the code trellis for the main decoder is biased, they eliminated those states whose occurring probabilities are nearly zero. (Hence, the scheme is called PSS (probability selecting states).) More precisely, from among $2^6\!=\!64$ states, $22$ states with lowest probabilities are eliminated for the above code. Then the number of states used for decoding is $42$ and $42$ survivors are preserved. Computer simulations show that the performance of a PSS-type decoder is as good as that of the conventional Viterbi decoder, whereas the hardware complexity of the former decoder is almost $1/2$ of that of the latter one.

\subsection{Trellis Degeneration Using Zero-Strings}
There exists a method where biased state distributions are indirectly used for complexity reduction in the main decoder. First consider an error trellis. Given a received data $\mbox{\boldmath $z$}\!=\!\{\mbox{\boldmath $z$}_k\}$, let $T_e$ be the corresponding error trellis. Note that unlike the code trellis $T_c$, the paths through $T_e$ have different a priori probabilities in general. Consequently, when $T_e$ is constructed based on the syndrome $\mbox{\boldmath $\zeta$}\!=\!\{\mbox{\boldmath $\zeta$}_k\}$ (which is computed using $\mbox{\boldmath $z$}^h\!=\!\{\mbox{\boldmath $z$}_k^h\}$), $T_e$ usually has many redundant paths that can be deleted in advance. Using this fundamental property of error trellises, Ariel and Snyders~\cite{arie 95,arie 99} proposed several methods to simplify $T_e$. Among them trellis degeneration using zero-strings~\cite{arie 95,arie 99} is most effective.
\par
In the following, $k_0$ is assumed to be $(n_0-1)$ for simplicity. Let $\mbox{\boldmath $\zeta$}\!=\!\{\zeta_k\}$ be the syndrome. An interval $[t, t']$ is called a {\it zero-string} if $\zeta_k\!=\!0,~t+1 \leq k \leq t'$. Note that within a zero-string, any two consecutive zero states (denoted {\boldmath $0$}) are connected by a zero-weight branch. Hence, if state {\boldmath $0$} has the least weight at $s \in [t, t']$, then state {\boldmath $0$} continues to have the least weight in $[s+1, t']$. We remark that this principle also holds in the reverse direction. Here suppose that we can identify a sub-interval $[\tau, \tau']$ of $[t, t']$ such that the all-zero path connecting state {\boldmath $0$} at depth $\tau$ and state {\boldmath $0$} at depth $\tau'$ is a portion of the overall ML path. In this case, all but the all-zero path connecting those states can be deleted. That is, $T_e$ is simplified in the interval $[\tau, \tau']$. This procedure is called {\it trellis degeneration}~\cite{arie 95,arie 99}.
\par
On the other hand, we already know that SST Viterbi decoding based on a code trellis and syndrome decoding based on the corresponding error trellis are equivalent. Hence, it is reasonable to think that trellis degeneration is equally possible in the code trellis for the main decoder in an SST Viterbi decoder~\cite{taji 971,taji 972}.
\par
{\it Remark 1:} The following argument is almost the same as that in~\cite{taji 972}. Also, the material is taken from it. To the best of our knowledge, however, when the work of~\cite{taji 972} was published (1997), the equivalence between SST Viterbi decoding based on a code trellis and syndrome decoding based on the corresponding error trellis had not been obtained. On the other hand, since the equivalence between the two schemes has been shown by now, the results about an error trellis can be transformed to the associated code trellis for the main decoder. That is, the application of the results in~\cite{arie 95,arie 99} to the code trellis for the main decoder is justified.
\par
First (see Section II) note that the hard-decision input to the main decoder is given by
\begin{displaymath}
\mbox{\boldmath $r$}_k^h=\zeta_k(H^{-1})^T .
\end{displaymath}
Also, in the case of QLI codes, the hard-decision input to the main decoder becomes
\begin{displaymath}
\mbox{\boldmath $\eta$}_{k-L}^h=(\zeta_k, \zeta_k) .
\end{displaymath}
Hence, an interval with $\zeta_k\!=\!0$ is transformed to an interval with $\mbox{\boldmath $r$}_k^h\!=\!\mbox{\boldmath $0$}$ (or $\mbox{\boldmath $\eta$}_{k-L}^h\!=\!\mbox{\boldmath $0$}$). In this paper, we call the latter (i.e., an interval where the hard-decision input to the main decoder is consecutively zero) a {\it zero-string} as well. We describe the trellis degeneration in the code trellis for the main decoder in more detail.
\par
{\it Code trellis degeneration using zero-strings~\cite{taji 972}:}
\begin{itemize}
\item[1)] Given a zero-string $[t, t']$, decode forward the code trellis from state $\mbox{\boldmath $x$}~(\neq \mbox{\boldmath $0$})$ at depth $t$. Let $\tau(\mbox{\boldmath $x$})$ be the first depth at which the metric of state {\boldmath $0$} is largest.
\item[2)] Similarly, decode backward the code trellis from state $\mbox{\boldmath $x$}'~(\neq \mbox{\boldmath $0$})$ at depth $t'$. Let $\tau'(\mbox{\boldmath $x$}')$ be the first depth at which the metric of state {\boldmath $0$} is largest.
\item[3)] Let $\tau\stackrel{\triangle}{=}\max_{\mbox{\boldmath $x$}}\tau(\mbox{\boldmath $x$})$. Also, let $\tau'\stackrel{\triangle}{=}\min_{\mbox{\boldmath $x$}'}\tau'(\mbox{\boldmath $x$}')$. If $\tau,~\tau' \in [t, t']$ and $\tau < \tau'$, then delete all the sub-paths in $[\tau, \tau']$ except for the all-zero sub-path. (That is, the code trellis is simplified in the interval $[\tau, \tau']$. In this case, we call trellis degeneration ``successful''.)
\end{itemize}
\par
{\it Remark 2:} The starting depths of the forward and the backward decoding can be chosen as $\tilde t(\leq t)$ and $\tilde t'(\geq t')$, respectively.
\par
{\it Remark 3:} Suppose that the length of a zero-string $[t, t']$ (denoted by $\ell$) has an appropriate value. Then for hard-decision data, the length $\ell_H\stackrel{\triangle}{=}(\tau-t)\!+\!(t'-\tau')$ can be determined in advance. Hence, for hard-decision data, if $\ell>\ell_H$ holds, then trellis degeneration is successful. For example, consider the code defined by $G\!=\!(1\!+\!D\!+\!D^2, 1\!+\!D^2)$. We have $\ell_H\!=\!(\tau-t)\!+\!(t'-\tau')\!=\!5\!+\!5\!=\!10$.
\par
Next, evaluate the complexity of Viterbi decoding where the trellis degeneration procedure is employed. Since trellis degeneration is rather complicated in a general case, we apply the procedure to those zero-strings whose lengths are larger than or equal to $\ell_0$, where $\ell_0$ is a predetermined value. Let $[t_j, t_j']$ be any such zero-string ($j$ is used to distinguish zero-strings). It is assumed that trellis degeneration is successful for each $[t_j, t_j']$. Let $N_s$ be the number of states in the trellis. Also, let $M$ be the section length of the trellis. We regard the computational complexity needed to decode one trellis section as one unit. (Then the Viterbi decoding complexity required to decode the whole trellis is given by $M$.) Under these conditions, let us evaluate the complexity of Viterbi decoding. Since trellis degeneration is successful for each zero-string $[t_j, t_j']$, the decoding complexity is reduced by
\begin{equation}
\Delta\stackrel{\triangle}{=}\sum_j(\tau_j'-\tau_j)
\end{equation}
as compared with the conventional decoding. On the other hand, in order to identify the sub-interval $[\tau_j, \tau_j']$ of $[t_j, t_j']$, the forward and the backward decoding are performed while changing the starting state. Let $\Delta'$ be the required computational complexity. Then the decoding complexity increases by
\begin{eqnarray}
\Delta' &\approx& \sum_j \bigl((N_s-1) \!\times\! (\tau_j-t_j)\!+\!(N_s-1) \!\times\! (t_j'-\tau_j')\bigr) \nonumber \\
&=& \sum_j(N_s-1)\bigl((\tau_j-t_j)+(t_j'-\tau_j')\bigr) .
\end{eqnarray}
Therefore, the overall decoding complexity is estimated as
\begin{equation}
Q_c \approx M+\Delta'-\Delta .
\end{equation}
Hence, if $\Delta'<\Delta$, then complexity reduction is realized. In particular, if
\begin{equation}
(N_s-1)\bigl((\tau_j-t_j)+(t_j'-\tau_j')\bigr)<\tau_j'-\tau_j
\end{equation}
holds for each $j$, then we have $\Delta'<\Delta$. Here note that for hard-decision data, we have
\begin{displaymath}
(\tau_j-t_j)+(t_j'-\tau_j')=\ell_H .
\end{displaymath}
Hence, if
\begin{displaymath}
(N_s-1)\ell_H<\tau_j'-\tau_j
\end{displaymath}
holds approximately, then we can expect to have $\Delta'<\Delta$. In this case, the length of the corresponding zero-string (i.e., $\ell$) becomes
\begin{displaymath}
\ell \approx (\tau_j'-\tau_j)+\ell_H .
\end{displaymath}
That is, if the condition
\begin{displaymath}
(N_s-1)\ell_H+\ell_H < (\tau_j'-\tau_j)+\ell_H ,
\end{displaymath}
i.e., 
\begin{equation}
N_s \times \ell_H < \ell
\end{equation}
holds, then complexity reduction is expected to occur. We can use the above inequality as a criterion for the length of a zero-string required for complexity reduction.
\par
{\it Example 3~\cite{taji 972}:} In connection with the above subject, computer simulations have been done using the QLI code $C_1$ defined in Example 1, where $M\!=\!10^5$ and $8$-level receiver quantization is assumed. Under these conditions, the behavior of the main decoder was investigated. Table VII shows the number of zero-strings whose lengths are larger than or equal to $\ell_0$. Table VIII shows the average length of zero-strings counted in Table VII. We observe that as the SNR increases, the zero-strings become less numerous and longer.
\begin{table}[tb]
\caption{Number of zero-strings}
\label{Table 7}
\begin{center}
\begin{tabular}{c*{5}{|c}}
$E_b/N_0\,(\mbox{dB})$ & $\ell_0\!=\!10$ & $\ell_0\!=\!15$ & $\ell_0\!=\!20$ & $\ell_0\!=\!25$ & $\ell_0\!=\!30$ \\
\hline
$4$ & $2761$ & $1527$ & $879$ & $490$ & $287$ \\
$5$ & $2948$ & $2003$ & $1373$ & $936$ & $651$ \\
$6$ & $2602$ & $2056$ & $1634$ & $1290$ & $1040$ \\
$7$ & $1808$ & $1590$ & $1398$ & $1236$ & $1080$ \\
$8$ & $1006$ & $953$ & $907$ & $851$ & $792$ \\
$9$ & $427$ & $425$ & $415$ & $407$ & $395$ \\
$10$ & $148$ & $148$ & $145$ & $145$ & $144$ 
\end{tabular}
\end{center}
\end{table}
\begin{table}[tb]
\caption{Average length of zero-strings}
\label{Table 8}
\begin{center}
\begin{tabular}{c*{5}{|c}}
$E_b/N_0\,(\mbox{dB})$ & $\ell_0\!=\!10$ & $\ell_0\!=\!15$ & $\ell_0\!=\!20$ & $\ell_0\!=\!25$ & $\ell_0\!=\!30$ \\
\hline
$4$ & $18.1$ & $23.3$ & $28.1$ & $33.1$ & $37.7$ \\
$5$ & $22.6$ & $27.7$ & $32.7$ & $37.8$ & $42.6$ \\
$6$ & $31.2$ & $36.4$ & $41.4$ & $46.6$ & $51.3$ \\
$7$ & $50.3$ & $55.5$ & $60.8$ & $65.9$ & $71.5$ \\
$8$ & $95.4$ & $100.1$ & $104.3$ & $109.7$ & $115.8$ \\
$9$ & $230.7$ & $231.7$ & $236.9$ & $241.1$ & $247.6$ \\
$10$ & $672.4$ & $672.4$ & $685.9$ & $685.9$ & $685.9$ 
\end{tabular}
\end{center}
\end{table}
\par
The normalized decoding complexity $Q_c/M$ obtained from simulations is given in Table IX. Since trellis degeneration is successful for almost all zero-strings of length $\ell \geq 15$, $\ell_0 \geq 20$ is assumed. In this example, the starting depths of the forward and the backward decoding for a zero-string $[t, t']$ are chosen as $t\!-\!1$ and $t'\!+\!1$, respectively.
\par
Now evaluate the length of a zero-string required for complexity reduction. Taking into account the starting depths of the forward and the backward decoding, we have
\begin{displaymath}
(N_s-1)\bigl((\tau_j-t_j)\!+\!(t_j'-\tau_j')\!+\!2 \bigr) < \tau_j'-\tau_j .
\end{displaymath}
\begin{displaymath}
\bigl(\mbox{i.e.,}~(N_s-1)(\ell_H\!+\!2)+\ell_H < (\tau_j'-\tau_j)+\ell_H . \bigr)
\end{displaymath}
Note that $N_s\!=\!4$ and $\ell_H\!=\!10$. Hence, if
\begin{equation}
46 < \ell
\end{equation}
holds, then we can expect that complexity reduction is realized. Accordingly, using Table VIII, let us search for the SNR at which the average length of zero-strings is nearly equal to $46$. We see that this value is attained at an SNR of $E_b/N_0\!=\!6\!\sim\!7 \mbox{dB}$ for $\ell_0\!=\!20$. Similarly, we see that $E_b/N_0 \!\approx\! 6 \mbox{dB}$ for $\ell_0\!=\!25$ and $E_b/N_0\!=\!5\!\sim\!6 \mbox{dB}$ for $\ell_0\!=\!30$. From Table IX, it is confirmed that these values are almost equal to the SNRs at which the decoding complexity is less than $1$ for the first time. Hence, the derived criterion for complexity reduction (i.e., $N_s \times \ell_H<\ell$) seems to be reasonable.
\begin{table}[tb]
\caption{Normalized decoding complexity}
\label{Table 9}
\begin{center}
\begin{tabular}{c*{3}{|c}}
$E_b/N_0\,(\mbox{dB})$ & $\ell_0\!=\!20$ & $\ell_0\!=\!25$ & $\ell_0\!=\!30$ \\
\hline
$4$ & $1.22$ & $1.10$ & $1.04$ \\
$5$ & $1.25$ & $1.12$ & $1.05$ \\
$6$ & $1.11$ & $1.02$ & $0.97$ \\
$7$ & $0.79$ & $0.76$ & $0.73$ \\
$8$ & $0.45$ & $0.44$ & $0.43$ \\
$9$ & $0.18$ & $0.18$ & $0.18$ \\
$10$ & $0.06$ & $0.06$ & $0.04$ 
\end{tabular}
\end{center}
\end{table}
\par
We remark that the derived criterion can be loosened. Note that for a trellis with large $N_s$, the condition seems to be strict. On the other hand, we already know that the state distribution in the main decoder is much biased under moderately noisy conditions. For example, consider the code trellis associated with the QLI code $C_2$ (cf. (85)). Here note the all-zero state and the states containing only one ``$1$'' (e.g., $(000001)$). We examined the total probability of these $7$ states. As a result (cf.~\cite{taji 86}), we have $87\%$ at $E_b/N_0\!=\!4\mbox{dB}$, $94\%$ at $E_b/N_0\!=\!5\mbox{dB}$, and $97\%$ at $E_b/N_0\!=\!6\mbox{dB}$. Hence, in order to identify the sub-interval $[\tau_j, \tau_j']$ of a zero-string $[t_j, t_j']$, we need not use all states ($\neq \mbox{\boldmath $0$}$) in the trellis as the starting state. That is, we can restrict the starting state to those $6$ states (the all-zero state is not used) under low to moderate noise level within a very small degradation. In this way, $N_s$ can be replaced by some smaller number. In this case, the values of $\tau_j$ and $\tau_j'$ may be slightly changed. A modified inequality can ease the criterion for complexity reduction.


\section{An Innovations Approach to ML Decoding of Block Codes}
In Section II, we have introduced the notion of innovations for Viterbi decoding of convolutional codes. The derived innovation is closely related to an SST Viterbi decoder which consists of a pre-decoder and a main decoder. The fundamental feature of the SST scheme lies in its structure where an {\it estimation error} is decoded in the main decoder. Here we see that a similar scheme (i.e., two-stage decoding) can be applied to block codes as well. Then it is reasonable to think that a kind of innovation can also be extracted in connection with ML decoding of block codes~\cite{lin 04}. In the following, we will show that this is actually possible.

\subsection{Two-Stage ML Decoding}
Let $G$ be a generator matrix for an $(n, k)$ block code, where its rank is assumed to be $k$. Let $H$ be a corresponding check matrix, where its rank is assumed to be $(n-k)$. Denote by $\mbox{\boldmath $i$}\!=\!\{i_j\}_{j=1}^k$ and $\mbox{\boldmath $i$}G\!=\!\mbox{\boldmath $y$}\!=\!\{y_j\}_{j=1}^n$ a message and the corresponding codeword, respectively. Here consider a two-stage ML decoding algorithm.
\par
i) First stage: Let $\mbox{\boldmath $z$}\!=\!\{z_j\}_{j=1}^n$ be a received data. The hard-decision received data is expressed as
\begin{equation}
\mbox{\boldmath $z$}^h=\mbox{\boldmath $y$}+\mbox{\boldmath $e$}=\mbox{\boldmath $i$}G+\mbox{\boldmath $e$} ,
\end{equation}
where $\mbox{\boldmath $e$}\!=\!\{e_j\}_{j=1}^n$ is an error. The transmitted message is estimated by using the inverse encoder $G^{-1}$. We have
\begin{equation}
\mbox{\boldmath $z$}^hG^{-1}=\mbox{\boldmath $i$}+\mbox{\boldmath $e$}G^{-1} .
\end{equation}
\par
ii) Second stage: The estimated message is re-encoded by $G$ and then the re-encoded data is added to the original received data {\boldmath $z$}. Let $\mbox{\boldmath $\xi$}\!=\!\{\xi_j\}_{j=1}^n$ be the result. We have
\begin{equation}
\mbox{\boldmath $\xi$}^h=\mbox{\boldmath $z$}^h+(\mbox{\boldmath $z$}^hG^{-1})G
\end{equation}
\begin{equation}
\xi_j=\left\{
\begin{array}{rl}
\vert z_j \vert,& \quad \xi_j^h=0 \\
- \vert z_j \vert,& \quad \xi_j^h=1 .
\end{array} \right.
\end{equation}
At the second stage, ML decoding is performed by regarding {\boldmath $\xi$} as a received data. Note that $\mbox{\boldmath $\xi$}^h$ is expressed as
\begin{eqnarray}
\mbox{\boldmath $\xi$}^h &=& (\mbox{\boldmath $i$}G+\mbox{\boldmath $e$})+(\mbox{\boldmath $i$}+\mbox{\boldmath $e$}G^{-1})G \nonumber \\
&=& (\mbox{\boldmath $e$}G^{-1})G+\mbox{\boldmath $e$} \\
&=& \mbox{\boldmath $u$}G+\mbox{\boldmath $e$} ,
\end{eqnarray}
where $\mbox{\boldmath $u$}\stackrel{\triangle}{=}\mbox{\boldmath $e$}G^{-1}$ is a message for the second-stage decoder and $\mbox{\boldmath $u$}G$ is the corresponding codeword. Hence, $\mbox{\boldmath $u$}\!=\!\mbox{\boldmath $e$}G^{-1}$ is decoded by the second-stage ML decoder. Finally, two decoder outputs are combined to produce the final decoder output, i.e.,
\begin{displaymath}
(\mbox{\boldmath $i$}+\mbox{\boldmath $u$})+\mbox{\boldmath $u$}=\mbox{\boldmath $i$} .
\end{displaymath}
\par
On the other hand, $\mbox{\boldmath $\xi$}^h$ has another expression. Since the rank of $G$ is $k$, $G$ can be decomposed as
\begin{equation}
G=A \times \Gamma \times B ,
\end{equation}
where $A\!=\!I_k$, $\Gamma\!=\!\left(
\begin{array}{cc}
I_k & O_{k, n-k} 
\end{array}
\right)$, and $B$ is an $n \times n$ non-singular matrix. Here the first $k$ rows of $B$ are equal to $G$ and the last $(n-k)$ columns of $B^{-1}$ are equal to $H^T$. As a result, we have
\begin{eqnarray}
I_n &=& B^{-1}B \nonumber \\
&=& \left(
\begin{array}{cc}
G^{-1} & H^T 
\end{array}
\right)\left(
\begin{array}{c}
G \\
(H^{-1})^T
\end{array}
\right) \nonumber \\
&=& G^{-1}G+H^T(H^{-1})^T .
\end{eqnarray}
Then
\begin{eqnarray}
\mbox{\boldmath $\xi$}^h &=& \mbox{\boldmath $e$}(G^{-1}G+I_n) \nonumber \\
&=& \mbox{\boldmath $e$}H^T(H^{-1})^T=\mbox{\boldmath $\zeta$}(H^{-1})^T
\end{eqnarray}
is obtained, where $\mbox{\boldmath $\zeta$}\!=\!\mbox{\boldmath $z$}^hH^T\!=\!\mbox{\boldmath $e$}H^T$ is the syndrome.
\par
In particular, let $G\!=\!\left(
\begin{array}{cc}
I_k & S 
\end{array}
\right)$, where $S$ is a $k \times (n-k)$ matrix. In this case, since
\begin{displaymath}
(H^{-1})^T=\left(
\begin{array}{cc}
O_{n-k, k} & I_{n-k} 
\end{array}
\right) ,
\end{displaymath}
we have
\begin{eqnarray}
\mbox{\boldmath $\xi$}^h &=& \mbox{\boldmath $\zeta$}(H^{-1})^T \nonumber \\
&=& \mbox{\boldmath $\zeta$}\left(
\begin{array}{cc}
O_{n-k, k} & I_{n-k} 
\end{array}
\right) \nonumber \\
&=& (O_{1, k}, \mbox{\boldmath $\zeta$}) .
\end{eqnarray}

\subsection{Innovations Associated with the Received Data for an ML Decoder}
The proposed two-stage ML decoding of block codes can also be discussed from an innovation viewpoint. In fact, the following argument is almost the same as that in Section II-A.
\par
Let
\begin{displaymath}
\mbox{\boldmath $z$}^h=\mbox{\boldmath $i$}G+\mbox{\boldmath $e$}
\end{displaymath}
be the hard-decision received data. By comparison with the linear filtering theory, consider the quantity
\begin{eqnarray}
\mbox{\boldmath $r$}^h &=& \mbox{\boldmath $z$}^h-\mbox{\boldmath $\hat i$}G \nonumber \\
&=& \mbox{\boldmath $z$}^h+\mbox{\boldmath $\hat i$}G ,
\end{eqnarray}
where $\mbox{\boldmath $\hat i$}$ denotes an estimate of {\boldmath $i$} based on $\mbox{\boldmath $z$}^h$. Suppose that $\mbox{\boldmath $\hat i$}$ has the form
\begin{equation}
\mbox{\boldmath $\hat i$}=\mbox{\boldmath $z$}^hP ,
\end{equation}
where $P$ is an $n \times k$ matrix. Then we have
\begin{eqnarray}
\mbox{\boldmath $r$}^h &=& \mbox{\boldmath $z$}^h+\mbox{\boldmath $z$}^hPG \nonumber \\
&=& (\mbox{\boldmath $i$}G+\mbox{\boldmath $e$})+(\mbox{\boldmath $i$}G+\mbox{\boldmath $e$})PG \nonumber \\
&=& \mbox{\boldmath $i$}(G+GPG)+\mbox{\boldmath $e$}PG+\mbox{\boldmath $e$} . \nonumber
\end{eqnarray}
Note that if
\begin{displaymath}
G+GPG=0
\end{displaymath}
or
\begin{displaymath}
GPG=G
\end{displaymath}
holds, then $\mbox{\boldmath $r$}^h$ is independent of {\boldmath $i$}. Here $GPG\!=\!G$ implies that $P$ is a {\it generalized inverse}~\cite{rao 72} of $G$. Then a right inverse $G^{-1}$ can be taken as $P$. In this case, $\mbox{\boldmath $r$}^h$ is independent of {\boldmath $i$} and we have
\begin{eqnarray}
\mbox{\boldmath $r$}^h &=& (\mbox{\boldmath $e$}G^{-1})G+\mbox{\boldmath $e$} \\
&=& \mbox{\boldmath $u$}G+\mbox{\boldmath $e$} \\
&=& \mbox{\boldmath $e$}(G^{-1}G+I_n) ,
\end{eqnarray}
where $\mbox{\boldmath $u$}\stackrel{\triangle}{=}\mbox{\boldmath $e$}G^{-1}$. We think this quantity corresponds to an innovation in the linear filtering theory. We remark that the right-hand side is just the input to the second-stage decoder in a two-stage ML decoder. Also, note that
\begin{eqnarray}
\mbox{\boldmath $r$}^hH^T &=& \mbox{\boldmath $z$}^hH^T+\mbox{\boldmath $z$}^hPGH^T \nonumber \\
&=& \mbox{\boldmath $z$}^hH^T=\mbox{\boldmath $\zeta$}
\end{eqnarray}
holds irrespective of $P$, where {\boldmath $\zeta$} is the syndrome. Hence, $\mbox{\boldmath $r$}^h$ and $\mbox{\boldmath $z$}^h$ generate the same syndrome {\boldmath $\zeta$}.
\par
On the other hand, $\mbox{\boldmath $r$}^h$ has another expression, i.e.,
\begin{eqnarray}
\mbox{\boldmath $r$}^h &=& \mbox{\boldmath $e$}(G^{-1}G+I_n) \nonumber \\
&=& \mbox{\boldmath $e$}H^T(H^{-1})^T=\mbox{\boldmath $\zeta$}(H^{-1})^T .
\end{eqnarray}
\par
Therefore,  with respect to $\mbox{\boldmath $r$}^h$, we have the following:
\begin{itemize}
\item[1)] $\mbox{\boldmath $r$}^h\!=\!\mbox{\boldmath $e$}(G^{-1}G\!+\!I_n)$ holds and there is a correspondence between {\boldmath $e$} and $\mbox{\boldmath $r$}^h$ in the sense that they generate the same syndrome {\boldmath $\zeta$}.
\item[2)] $\mbox{\boldmath $r$}^h$ and $\mbox{\boldmath $z$}^h$ generate the same syndrome {\boldmath $\zeta$}.
\end{itemize}
These properties imply that we can regard $\mbox{\boldmath $r$}^h$ as the innovation corresponding to $\mbox{\boldmath $z$}^h$. We remark that the variable which represents time (or depth) is not assumed explicitly in block codes. That is, a codeword may not be regarded as a time series. Hence, we may call the extracted quantity the innovation in a {\it weak sense}~\cite{kuni 76}.
\par
Moreover, consider the mapping: $\mbox{\boldmath $z$}^h \!\mapsto\! \mbox{\boldmath $r$}^h\!=\!\mbox{\boldmath $z$}^h(G^{-1}G\!+\!I_n)$. It is shown that it is not invertible and the innovation $\mbox{\boldmath $r$}^h$ corresponding to $\mbox{\boldmath $z$}^h$ cannot be further reduced. Proofs are almost the same as those given in Section II-A.

\section{Conclusion}
In this paper, by comparing the results in the linear filtering theory, we have introduced the notion of innovations for Viterbi decoding of convolutional codes. It has been shown that the newly defined innovations are closely related to the structure of an SST Viterbi decoder. We have also shown that a similar result holds with respect to QLI codes. In this case, we have seen that the innovation-like quantity has a connection with linear smoothing of the information. Moreover, for a QLI code, we have clarified the relationship between the filtered estimate and the smoothed estimate of the information. We think the obtained results are due to having introduced innovations associated with the received data. With respect to innovations, it is written in~\cite{hida 11,hida 14} as follows:
\par Consider a complex system. Suppose that we have generated some simpler system composed of mutually independent elements. Also, suppose that for a given time $t$, the new system has the same information as the original one has by time $t$. Then the newly generated simpler system is called the innovations. It is not easy to obtain such an ideal system. For typical problems, however, the corresponding innovations have been derived. Obtaining innovations for a given complex system provides a method for the {\it reduction} of time series or stochastic processes.
\par In those books, the innovations method is regarded as an essentially important tool for {\it reduction} $\rightarrow$ {\it synthesis} $\rightarrow$ {\it analysis} of a given complex system. In our case, the known SST scheme has been more clarified using innovations. Furthermore, we have shown the proposed innovations approach can be extended to block codes as well. In fact, a kind of innovation has been extracted in connection with ML decoding of block codes.


%

\appendices
\section{Proof of Lemma 3.2}
\renewcommand{\theequation}{\thesection.\arabic{equation}}
\addtocounter{equation}{-108}
Without loss of generality, for
\begin{displaymath}
\alpha_1\stackrel{\triangle}{=}P(e_k^{(1)}\!=\!0, r_k^{(1)h}\!=\!1)+P(e_k^{(1)}\!=\!1, r_k^{(1)h}\!=\!0) ,
\end{displaymath}
we will show that $0 \leq \alpha_1 \leq 1/2$. In the following, we omit the delay operator $D$ for simplicity. Let
\begin{equation}
G=\left(
\begin{array}{cccc}
g_{1, 1} & g_{1, 2} & \ldots & g_{1, n_0}  \\
g_{2, 1} & g_{2, 2} & \ldots & g_{2, n_0}  \\
\ldots & \ldots & \ldots & \ldots \\
g_{k_0, 1} & g_{k_0, 2} & \ldots & g_{k_0, n_0}
\end{array}
\right)
\end{equation}
be the generator matrix. Also, let
\begin{equation}
G^{-1}=\left(
\begin{array}{cccc}
b_{1, 1} & b_{1, 2} & \ldots & b_{1, k_0}  \\
b_{2, 1} & b_{2, 2} & \ldots & b_{2, k_0}  \\
\ldots & \ldots & \ldots & \ldots \\
b_{n_0, 1} & b_{n_0, 2} & \ldots & b_{n_0, k_0}
\end{array}
\right)
\end{equation}
be a right inverse of $G$. Then the first column of
\begin{eqnarray}
G^{-1}G+I_{n_0} &=& \left(
\begin{array}{cccc}
b_{1, 1} & b_{1, 2} & \ldots & b_{1, k_0}  \\
b_{2, 1} & b_{2, 2} & \ldots & b_{2, k_0}  \\
\ldots & \ldots & \ldots & \ldots \\
b_{n_0, 1} & b_{n_0, 2} & \ldots & b_{n_0, k_0}
\end{array}
\right) \nonumber \\
&& \qquad \times \left(
\begin{array}{cccc}
g_{1, 1} & g_{1, 2} & \ldots & g_{1, n_0}  \\
g_{2, 1} & g_{2, 2} & \ldots & g_{2, n_0}  \\
\ldots & \ldots & \ldots & \ldots \\
g_{k_0, 1} & g_{k_0, 2} & \ldots & g_{k_0, n_0}
\end{array}
\right) \nonumber \\
&& +\left(
\begin{array}{cccc}
1 & 0 & \ldots & 0 \\
0 & 1 & \ldots & 0 \\
\ldots & \ldots & \ldots & \ldots \\
0 & 0 & \ldots & 1
\end{array}
\right) \nonumber
\end{eqnarray}
is given by
\begin{displaymath}
\left(
\begin{array}{c}
b_{1, 1}g_{1, 1}\!+\!b_{1, 2}g_{2, 1}\!+\! \cdots \!+\!b_{1, k_0}g_{k_0, 1}\!+\!1  \\
b_{2, 1}g_{1, 1}\!+\!b_{2, 2}g_{2, 1}\!+\! \cdots \!+\!b_{2, k_0}g_{k_0, 1} \\
\cdots \\
b_{n_0, 1}g_{1, 1}\!+\!b_{n_0, 2}g_{2, 1}\!+\! \cdots \!+\!b_{n_0, k_0}g_{k_0, 1}
\end{array}
\right) .
\end{displaymath}
Hence, it follows that
\begin{eqnarray}
r_k^{(1)h} &=& e_k^{(1)}(b_{1, 1}g_{1, 1}\!+\!b_{1, 2}g_{2, 1}\!+\! \cdots \!+\!b_{1, k_0}g_{k_0, 1}\!+\!1) \nonumber \\
&& +e_k^{(2)}(b_{2, 1}g_{1, 1}\!+\!b_{2, 2}g_{2, 1}\!+\! \cdots \!+\!b_{2, k_0}g_{k_0, 1}) \nonumber \\
&& \cdots \nonumber \\
&& +e_k^{(n_0)}(b_{n_0, 1}g_{1, 1}\!+\!b_{n_0, 2}g_{2, 1}\!+\! \cdots \!+\!b_{n_0, k_0}g_{k_0, 1}) \nonumber \\
&=& \tilde r_k^{(1)h}+e_k^{(1)} ,
\end{eqnarray}
where
\begin{eqnarray}
\tilde r_k^{(1)h} &\stackrel{\triangle}{=}& e_k^{(1)}(b_{1, 1}g_{1, 1}\!+\!b_{1, 2}g_{2, 1}\!+\! \cdots \!+\!b_{1, k_0}g_{k_0, 1}) \nonumber \\
&& +e_k^{(2)}(b_{2, 1}g_{1, 1}\!+\!b_{2, 2}g_{2, 1}\!+\! \cdots \!+\!b_{2, k_0}g_{k_0, 1}) \nonumber \\
&& \cdots \nonumber \\
&& +e_k^{(n_0)}(b_{n_0, 1}g_{1, 1}\!+\!b_{n_0, 2}g_{2, 1}\!+\! \cdots \nonumber \\
&& \qquad +b_{n_0, k_0}g_{k_0, 1}) .
\end{eqnarray}
\par
Here note the definition of $\alpha_1$.
\begin{itemize}
\item[1)] $e_k^{(1)}\!=\!0,r_k^{(1)h}\!=\!1$: This is equivalent to $e_k^{(1)}\!=\!0,\tilde r_k^{(1)h}\!=\!1$.
\item[2)] $e_k^{(1)}\!=\!1,r_k^{(1)h}\!=\!0$: This is equivalent to $e_k^{(1)}\!=\!1,\tilde r_k^{(1)h}\!=\!1$.
\end{itemize}
Hence, we have
\begin{eqnarray}
\alpha_1 &=& P(e_k^{(1)}\!=\!0, r_k^{(1)h}\!=\!1)+P(e_k^{(1)}\!=\!1, r_k^{(1)h}\!=\!0) \nonumber \\
&=& P(e_k^{(1)}\!=\!0, \tilde r_k^{(1)h}\!=\!1)+P(e_k^{(1)}\!=\!1, \tilde r_k^{(1)h}\!=\!1) \nonumber \\
&=& P(\tilde r_k^{(1)h}=1).
\end{eqnarray}
Since $\tilde r_k^{(1)h}$ is the sum of error terms, we can assume that $\alpha_1$ has the form
\begin{equation}
\alpha_1=P(e_1\!+\!e_2\!+\!\cdots \!+\!e_n\!=\!1) ,
\end{equation}
where errors $e_j$ are mutually independent. In the following, $n$ is assumed to be even without loss of generality.
\par
In order to evaluate the right-hand side, consider the binominal expansion:
\begin{eqnarray}
\lefteqn{\bigl((1-\epsilon)+\epsilon \bigr)^n} \nonumber \\
&& ={}_nC_0(1-\epsilon)^n+{}_nC_1\epsilon (1-\epsilon)^{n-1}+\cdots \nonumber \\
&& \qquad +{}_nC_{n-1}\epsilon^{n-1}(1-\epsilon)+{}_nC_n\epsilon^n \nonumber \\
&& =\bigl({}_nC_0(1-\epsilon)^n+{}_nC_2\epsilon^2(1-\epsilon)^{n-2}+\cdots \nonumber \\
&& \qquad +{}_nC_{n-2}\epsilon^{n-2}(1-\epsilon)^2+{}_nC_n\epsilon^n \bigr) \nonumber \\
&& ~~+\bigl({}_nC_1\epsilon(1-\epsilon)^{n-1}+{}_nC_3\epsilon^3(1-\epsilon)^{n-3}+\cdots \nonumber \\
&& \qquad +{}_nC_{n-1}\epsilon^{n-1}(1-\epsilon)\bigr) \nonumber \\
&& =h(\epsilon)+f(\epsilon) ,
\end{eqnarray}
where
\begin{eqnarray}
h(\epsilon) &\stackrel{\triangle}{=}& {}_nC_0(1-\epsilon)^n+{}_nC_2\epsilon^2(1-\epsilon)^{n-2}+\cdots \nonumber \\
&& \qquad +{}_nC_{n-2}\epsilon^{n-2}(1-\epsilon)^2+{}_nC_n\epsilon^n \\
f(\epsilon) &\stackrel{\triangle}{=}& {}_nC_1\epsilon(1-\epsilon)^{n-1}+{}_nC_3\epsilon^3(1-\epsilon)^{n-3}+\cdots \nonumber \\
&& \qquad +{}_nC_{n-1}\epsilon^{n-1}(1-\epsilon) .
\end{eqnarray}
Note that $\alpha_1\!=\!f(\epsilon)$. We will show the following:
\begin{itemize}
\item[1)] $f(0)=0$
\item[2)] $f(1/2)=1/2$
\item[3)] $f(\epsilon)$ is monotone increasing for $0 \leq \epsilon \leq 1/2$.
\end{itemize}
\par
1) is obvious. Let us show 2). Note that
\begin{eqnarray}
f(1/2) &=& {}_nC_1\left(\frac{1}{2}\right)^n\!+\!{}_nC_3\left(\frac{1}{2}\right)^n\!+\!\cdots \!+\!{}_nC_{n-1}\left(\frac{1}{2}\right)^n \nonumber \\
&=& \bigl({}_nC_1\!+\!{}_nC_3\!+\!\cdots \!+\!{}_nC_{n-1}\bigr)\left(\frac{1}{2}\right)^n \nonumber \\
&=& 2^{n-1} \times \left(\frac{1}{2}\right)^n=1/2 ,
\end{eqnarray}
where the equality ${}_nC_1\!+\!{}_nC_3\!+\!\cdots \!+\!{}_nC_{n-1}\!=\!2^{n-1}$~\cite{spi 07} is used.
\par
Finally, we will show 3). Since $h(\epsilon)\!+\!f(\epsilon)\!=\!1$,
\begin{displaymath}
h'(\epsilon)+f'(\epsilon)=0 
\end{displaymath}
holds (``$'$'' means differentiation with respect to $\epsilon$). Hence, $f'(\epsilon) \geq 0$ is equivalent to $h'(\epsilon) \leq 0$. Then we will show the latter for $0 \leq \epsilon \leq 1/2$. From the definition of $h(\epsilon)$, we have
\begin{eqnarray}
h'(\epsilon) &=& -n(1-\epsilon)^{n-1}+n(n-1)\epsilon(1-\epsilon)^{n-2} \nonumber \\
&& -\frac{n(n-1)(n-2)}{2 \times 1}\epsilon^2(1-\epsilon)^{n-3} \nonumber \\
&& +\cdots +\frac{n(n-1)(n-2)}{2 \times 1}\epsilon^{n-3}(1-\epsilon)^2 \nonumber \\
&& -n(n-1)\epsilon^{n-2}(1-\epsilon)+n\epsilon^{n-1} \nonumber \\
&=& (-n) \times \bigl((1-\epsilon)^{n-1}-(n-1)\epsilon(1-\epsilon)^{n-2} \nonumber \\
&& +\frac{(n-1)(n-2)}{2 \times 1}\epsilon^2(1-\epsilon)^{n-3} \nonumber \\
&& -\cdots -\frac{(n-1)(n-2)}{2 \times 1}\epsilon^{n-3}(1-\epsilon)^2 \nonumber \\
&& +(n-1)\epsilon^{n-2}(1-\epsilon)-\epsilon^{n-1} \bigr) \nonumber \\
&=& -n \bigl((1-\epsilon)-\epsilon \bigr)^{n-1} \nonumber \\
&=& -n(1-2\epsilon)^{n-1}\leq 0~(0 \leq \epsilon \leq 1/2) .
\end{eqnarray}
Thus 3) is proved. This completes the proof of the lemma.


\section{Proof of Lemma 3.4}
Without loss of generality, for
\begin{displaymath}
\beta_1\stackrel{\triangle}{=}P(e_{k-L}^{(1)}\!=\!0, \zeta_k\!=\!1)+P(e_{k-L}^{(1)}\!=\!1, \zeta_k\!=\!0) ,
\end{displaymath}
we will show that $0 \leq \beta_1 \leq 1/2$. Let
\begin{equation}
G=(g_1, g_2),~g_1+g_2=D^L
\end{equation}
be a generator matrix of a QLI code. Since the check matrix is given by $H\!=\!(g_2, g_1)$, we have
\begin{eqnarray}
\zeta_k &=& \mbox{\boldmath $e$}_kH^T=(e_k^{(1)}, e_k^{(2)})\left(
\begin{array}{c}
g_2
\\
g_1
\end{array}
\right) \nonumber \\
&=& e_k^{(1)}g_2+e_k^{(2)}g_1 . \nonumber
\end{eqnarray}
\par
First consider the case 1) $e_{k-L}^{(1)}\!=\!0,~\zeta_k\!=\!1$. Since $\zeta_k$ is rewritten as
\begin{eqnarray}
\zeta_k &=& e_k^{(1)}(g_1\!+\!g_2)+e_k^{(1)}g_1+e_k^{(2)}g_1 \nonumber \\
&=& e_{k-L}^{(1)}+e_k^{(1)}g_1+e_k^{(2)}g_1 , \nonumber
\end{eqnarray}
1) is equivalent to $e_{k-L}^{(1)}\!=\!0,~e_k^{(1)}g_1\!+\!e_k^{(2)}g_1\!=\!1$.
\par
Next, consider the case 2) $e_{k-L}^{(1)}\!=\!1,~\zeta_k\!=\!0$. We see that this is equivalent to $e_{k-L}^{(1)}\!=\!1,~e_k^{(1)}g_1\!+\!e_k^{(2)}g_1\!=\!1$. Hence, we have
\begin{eqnarray}
\beta_1 &=& P(e_{k-L}^{(1)}\!=\!0, \zeta_k\!=\!1)+P(e_{k-L}^{(1)}\!=\!1, \zeta_k\!=\!0) \nonumber \\
&=& P(e_{k-L}^{(1)}\!=\!0, e_k^{(1)}g_1\!+\!e_k^{(2)}g_1\!=\!1) \nonumber \\
&& \qquad +P(e_{k-L}^{(1)}\!=\!1, e_k^{(1)}g_1\!+\!e_k^{(2)}g_1\!=\!1) \nonumber \\
&=& P(e_k^{(1)}g_1+e_k^{(2)}g_1=1) .
\end{eqnarray}
As in the case of Lemma 3.2, the right-hand side is less than or equal to $1/2$ for $0 \leq \epsilon \leq 1/2$. This proves the lemma.



\section*{Acknowledgment}
The author is very grateful to the anonymous reviewers and the associate editor for their constructive comments and suggestions, which have greatly improved the presentation of this paper. He also thanks Dr. Shigeichi Hirasawa, a Professor Emeritus at Waseda University, for his support during the course of this work.

\ifCLASSOPTIONcaptionsoff
  \newpage
\fi

\begin{IEEEbiographynophoto}{Masato Tajima}
(M'86--SM'13) was born in Toyama, Japan, on August 13, 1949. He received the B.E., M.E., and Dr. of Eng. degrees all in electrical engineering from Waseda University, Tokyo, Japan, in 1972, 1974, and 1979, respectively. He joined the Electronics Equipment Laboratory of Toshiba R$\&$D Center in 1979, where he engaged in research and development of channel coding techniques with applications to satellite communication systems. From 1993 to 2006, he was with the Department of Intellectual Information Systems Engineering, Toyama University, first as an Associate Professor, next as a Professor. From 2006 to 2015, he was with the Graduate School of Science and Engineering, University of Toyama, as a Professor. He is currently a Professor Emeritus at University of Toyama. His research interests are in coding theory and its applications.
\end{IEEEbiographynophoto}






\end{document}